\documentclass[10pt,journal,compsoc]{IEEEtran}

\ifCLASSOPTIONcompsoc
  \usepackage[nocompress]{cite}
\else
  \usepackage{cite}
\fi

\usepackage{booktabs} 
\usepackage{caption} 
\usepackage{graphicx}
\usepackage{epstopdf}
\usepackage{xcolor}
\usepackage{url}
\usepackage{cite}
\usepackage{hyperref}
\usepackage{multirow}
\usepackage{adjustbox}
\usepackage{makecell}

\newcommand{\tcomp}{\textit{Tcomp}}
\newcommand{\tcomm}{\textit{Tcomm}}
\newcommand{\tcopy}{\textit{Tcopy}}
\newcommand{\tslack}{\textit{Tslack}}
\newcommand{\greta}{\textit{COUNTDOWN Slack}}
\newcommand{\fermata}{\textit{Fermata}}
\newcommand{\andante}{\textit{Andante}}
\newcommand{\adagio}{\textit{Adagio}}
\newcommand{\cntd}{\textit{COUNTDOWN}}

\newcommand{\plain}{\textit{Baseline}}

\newcommand{\mfreq}{\textit{Min Freq}}

\begin{document}

\title{COUNTDOWN Slack: a Run-time Library to Reduce Energy Footprint in Large-scale MPI Applications}

\author{Daniele~Cesarini,
        Andrea~Bartolini,~\IEEEmembership{Member,~IEEE,}
        Andrea~Borghesi,\\
        Carlo~Cavazzoni,
        Mathieu~Luisier,~\IEEEmembership{Member,~IEEE,}
        and~Luca~Benini,~\IEEEmembership{Fellow,~IEEE}


\IEEEcompsocitemizethanks{
    \IEEEcompsocthanksitem D. Cesarini and A. Bartolini are with the Department of Electrical, Electronic and Information Engineering "Guglielmo Marconi", University of Bologna, 40136 Bologna, Italy (e-mail: daniele.cesarini@unibo.it;a.bartolini@unibo.it).
    \IEEEcompsocthanksitem A. Borghesi is with the Department of Computer Science and Engineering, University of Bologna, 40136 Bologna, Italy (e-mail: andrea.borghesi3@unibo.it).
    \IEEEcompsocthanksitem C. Cavazzoni is with the Department of SuperComputing Applications and Innovation, CINECA, 40033 Casalecchio di Reno (BO), Italy (e-mail: c.cavazzoni@cineca.it).
    \IEEEcompsocthanksitem M.Luisier is with the Department of Information Technology and Electrical Engineering, Swiss Federal Institute of Technology in Zurich, 8092 Zurich, Switzerland (e-mail: mluisier@iis.ee.ethz.ch).
    \IEEEcompsocthanksitem L. Benini is with the Department of Information Technology and Electrical Engineering, Swiss Federal Institute of Technology in Zurich, 8092 Zurich, Switzerland and the Department of Electrical, Electronic and Information Engineering "Guglielmo Marconi", University of Bologna, 40136 Bologna, Italy (e-mail: lbenini@iis.ee.ethz.ch).}
}



\maketitle

\IEEEdisplaynontitleabstractindextext
\IEEEpeerreviewmaketitle

\begin{abstract}

The power consumption of supercomputers is a major challenge for system owners, users, and society. It limits the capacity of system installations, it requires large cooling infrastructures, and it is the cause of a large carbon footprint.
Reducing power during application execution without changing the application source code or increasing time-to-completion is highly desirable in real-life high-performance computing scenarios.

The power management run-time frameworks proposed in the last decade are based on the assumption that the duration of communication and application phases in an MPI application can be predicted and used at run-time to trade-off communication slack with power consumption. In this manuscript, we first show that this assumption is too general and leads to mispredictions, slowing down applications, thereby jeopardizing the claimed benefits. We then propose a new approach based on (i) the separation of communication phases and slack during MPI calls and (ii) a timeout algorithm to cope with the hardware power management latency, which jointly makes it possible to achieve performance-neutral power saving in MPI applications without requiring labor-intensive and risky application source code modifications. We validate our approach in a tier-1 production environment with widely adopted scientific applications. 
Our approach has a time-to-completion overhead lower than $1\%$, while it successfully exploits slack in communication phases to achieve an average energy saving of $10\%$.
If we focus on a large-scale application runs, the proposed approach achieves $22\%$ energy saving with an overhead of only $0.4\%$. With respect to state-of-the-art approaches, \greta{} is the only that always leads to an energy saving with negligible overhead ($<3\%$).

\end{abstract}

\begin{IEEEkeywords}
HPC, MPI, DVFS, power management, reactive policy.
\end{IEEEkeywords}

\maketitle

\section{Introduction}
\label{sec:introduction}


With the end of Dennard's scaling\cite{Dennards,DarkSilicon}, the last decade has seen a progressive increase of the power density required to operate each new processor generation at its maximum performance. 
Supercomputing installations have suffered from this power density increase, which over the years has pushed up the energy provisioning and cooling costs. While more efficient cooling techniques have been adopted to reduce the energy wasted at the infrastructure level, e.g. hot-water and free cooling \cite{LRZ,TII_conficoni,TSC_conficoni}, and more specialized computing elements with a higher ratio of vector and SIMD units with respect to general-purpose processors have emerged \cite{Top500,Green500}, but a lot remains to be done in practice to reduce the energy wasted during computation. 

Processor designers have addressed this aspect by embedding in their products finer and smarter power management support to automatically trade off performance for power consumption \cite{hackenberg2015energy,OCC_ISC17}. By mission and design, the high-performance and scientific computing sectors aim at maximizing the peak performance of the computing systems, hence these techniques are seen as detrimental to the time-to-solution and time-to-science, and often disabled \cite{cesarini2018energy}.

Indeed, low power design strategies enable computing resources to trade-off their performance for power consumption by means of low-power modes of operation. These power states are obtained by Dynamic and Voltage Frequency Scaling (DVFS) (also known as performance states or P-states \cite{ACPI}), clock gating or throttling states (T-states), and idle states which switch off unused resources (C-states \cite{ACPI}). While the built-in hardware and operating system (OS) policies are application-agnostic, in recent years several approaches have been proposed to let the final user control them in userspace \cite{fraternali_islped04,lrz_lowfreq,losalomos_sc05,GEOPM} and at execution time \cite{adagio_dynamic,Schulz_IPDPS10}.

The first family of approaches intends to trade-off power consumption and performance to gain energy efficiency \cite{lrz_lowfreq,fraternali_islped14,fraternali2018quantifying,losalomos_sc05}. These techniques explore the use of HW power management knobs and application parameters to study the execution time (Time-to-Solution, TtS), average power, and energy (energy-to-solution, EtS) dependency with respect to these knobs and parameters. While these approaches can be used in combination with autotuners and resource management frameworks to explore the EtS-TtS Pareto curve, they have a limited potential in the current supercomputing scenario: slowing down applications is almost always detrimental to the total cost of ownership (TCO) due to the large contribution related to the depreciation cost of the IT equipment \cite{borghesi2018pricing}. 

The second family of approaches focuses on improving application performance under a power cap \cite{borghesi2018scheduling,GEOPM,freeh2008just}. These approaches target power limited systems, computing nodes, and processing elements. They rely on the runtime capability of tracking the critical task in the application; then, the power budget of the node/socket/core running the critical task is dynamically relaxed while tightening the power budget of the non-critical resources. This not only involves software approaches \cite{borghesi2018scheduling}, but also HW power management solutions, like Intel Turbo mode, and RAPL\cite{rapl}. These methods are tailored to power capped supercomputing systems that still belong to a niche \cite{EEHPCJSRM}.

The third and last family of approaches aim at cutting the IT energy waste by reducing the performance of the processing elements when the application is in a phase with communication slack available \cite{adagio_static,adagio_dynamic,lim2006adaptive,kerbyson2011energy,freeh2008just,simil_adagio,MVAPICH2-EA,cesarini_andare_18,cesarini_countdown}. These approaches try to isolate at runtime regions of the application execution flow which can be executed at a reduced P-state without impacting the application performance. While the hardware power management logic in today's processing elements is effective in reducing the power consumption of idle resources, in large-scale MPI parallel applications that fully utilize all the assigned processing elements workload unbalance, synchronization, and communication slack can be exploited to save energy. Several works have been proposed to address this challenge. However, also for these approaches slowing down the application is detrimental for the TCO thus making performance-neutral approaches more appealing.

The power management run-time frameworks which have been proposed in the latter family are based on the assumption that the duration of communication and computation phases in an MPI application can be predicted at execution time. In this manuscript, we first show that this assumption is too optimistic and leads to mispredictions, slowing down the application execution time, which jeopardizes their benefits. We then propose \greta{}\footnote{Github Repository: \url{https://github.com/EEESlab/countdown}} a new approach based on (i) the separation of communication phases and slack during MPI calls and (ii) a timeout algorithm to cope with the hardware power management latency, which jointly allows us to achieve performance-neutral power saving in MPI applications. We validate our approach in a tier-1 production environment with a widely adopted scientific benchmark suite \cite{nas}, and a two-times ACM Gordon Bell Prize finalist application \cite{Luisier_3,Luisier_4}. We also compared \greta{} with the main state-of-the-art approaches. 
In average \greta{} reduces the energy consumption of $9.96\%$ with an average overhead of $0.79\%$.
When compared with the state of the art, \greta{} is capable of achieving similar energy saving but with negligible impact on the application performance. If we consider the worst-case performance degradation, \greta{} has a minimal impact on performance, just $3.02\%$ overhead, while the worst-case overhead for state-of-the-art approaches is between $8.92\%$ and $144.75\%$. If we consider that only a negligible overhead (below $5\%$) is acceptable, \greta{} is the only approach that never exceeds this value for all the applications while at the same time always leading to an energy saving. In contrast, state-of-the-art approaches can cause non-negligible overheads or severe energy losses. Worst-case energy saving for \greta{} is $1\%$, while for the state-of-the-art approaches it ranges from $0.05\%$ to $-24.69\%$.


The paper is organized as follows. Section 2, presents the state of the art in power and energy management approaches for HPC computing systems. Section 3 introduces a background of power-saving in MPI-based applications. Section 4 describes the implementation of our \greta{} runtime. Section 5 explains our implementation of the state of the art of the energy-aware runtime that we use to compare with \greta. In Section 6, we report an analysis of our benchmarks in term of predictability of computation and communication region of the application. Moreover, we report experimental results in terms of overhead, energy and power saving for production applications in a tier-1 supercomputer.

\section{Related Work}
\label{sec:related}

Energy Efficiency is a hot topic in supercomputing. 
In the last decade, several works were proposed to reduce the energy waste of large scale MPI applications.

Kappiah et al. \cite{freeh2008just} show that in an MPI parallel application it is possible to use the PMPI profiling interface \cite{PMPI} to intercept MPI calls and isolate the time spent by each rank during an application iteration in communication and computation. The authors show that in collective MPI primitives the amount of synchronization slack can be converted into power (and energy) reduction by slowing down the computation (lowering the processor's P-state) to absorb the available slack. This operation can be done under the assumption that the communication and computation time of the upcoming application iteration can be known upfront. The authors of the paper propose to use an error signal (desired vs measured slack) computed on the previous iteration to drive the P-state reduction. Iterations need to be marked by the user.

Lim et at. \cite{lim2006adaptive} show that during the communication time (time spent in the MPI library) of an MPI parallel application the cores' P-state can be significantly reduced without causing severe overheads; thus they propose to execute the communication phases at a reduced P-state and computing phases at the default one. Due to the latency of P-state transitions, it is not feasible to target all the communication phases singularly, but the authors propose to group them based on a proximity index and allow a P-state reduction only in regions of the code with higher communication density. Regions of the code are constructed and marked at execution time by leveraging the hash of the call stack when MPI primitives are encountered. The proposed algorithm uses the last value prediction to determine the beginning and the end of a given region or the P-state to be applied for the upcoming region. The P-state is selected based on the measured IPS (instructions per second) on the previous region and a pre-characterization of the optimal P-state for a given IPS level.

Sundriyal et al. \cite{MVAPICH2_PSTATE,MVAPICH2_ALL,MVAPICH2_Gather} focus on the All-to-All \cite{MVAPICH2_ALL}, send/receive \cite{MVAPICH2_PSTATE}, and AllGather communications \cite{MVAPICH2_Gather}. They analyze the impact of fine-grain power management strategies in MVAPICH2 communication primitives (considered singularly) and their results suggest that different regions in the MPI primitives have different power/performance trade-offs.

Rountree et al. \cite{adagio_static} propose to separate the communication time into the slack and copy time. The slack time is caused by waiting for the critical task to enter the MPI primitive, and the data transfer causes the copy time. The authors define the task as the region of code between two MPI communication calls and define an optimization problem to minimize the slack time and save power (and energy) by reducing the P-state of the core during computation time. In \cite{adagio_dynamic} the same authors propose three online algorithms (Fermata, Andante, and Adagio) that use, similarly to \cite{lim2006adaptive}, the hash of the call stack at the entrance of an MPI call as a TaskId to identify similar tasks. The Andante algorithm uses the last value prediction on previously executed tasks that have the same TaskId of the upcoming task to estimate the communication, slack time and IPS and select for the upcoming task the P-state which minimizes its predicted slack. Due to the finite numbers of P-state available, it is not always possible to nullify the slack time.
For this reason, the Fermata algorithm uses the last value prediction to estimate the remaining slack time of the upcoming task, and if it is expected to be twice larger than an empirical switching time threshold (100ms) the region is considered for slack reclamation. Only, in this case, a timer is set to expire after the switching time threshold, and in the call back the minimum P-state is applied. If the task (MPI call) terminates before the timer expires, the callback is canceled. Adagio combines Andante and Fermata. It must be noted that Fermata will potentially lower the P-state also during copy time and similarly to Andante can lead to misprediction and costly performance overhead or loss of energy-saving opportunities, which can become severe in irregular applications \cite{kerbyson2011energy}.

Bhalachandra et al. \cite{simil_adagio}, as in \cite{adagio_static,adagio_dynamic}, focus on saving power by entering a low power state for processes which are not in the critical path. The authors propose an algorithm to save energy by reducing application unbalance. This is based on measuring the start and end time of each MPI\_barrier and MPI\_Allreduce primitives to compute the duration of application and MPI code. Based on that, the authors propose a feedback loop to lower the P-state and T-state if in previous computation and MPI regions the overhead was below a given threshold. This algorithm is based on the assumption that the duration of the current application and MPI phases will be the same as the previous ones. 

Venkatesh et al. \cite{MVAPICH2-EA} show that the approaches based on temporal execution patterns for predicting slack (such as last value prediction) \cite{adagio_static,adagio_dynamic,lim2006adaptive,li2004thrifty} can lead to significant misprediction errors. The authors propose to use a combination of empirical observation and communication models specialized for the different classes of communication primitives for estimating the duration of the MPI phases. If this estimation is long enough, they will decide to reduce the P-state.

Cesarini et al. \cite{cesarini_andare_18,cesarini_countdown} tries to overcome mispredictions by leveraging a timeout policy (namely COUNTDOWN) which sets a timer at the entrance of any MPI calls (Fermata was doing it only in the one predicted to be long enough) to discard communication times shorter than the hardware power controller latency, which is measured to be 500$\mu$s, in line with the findings of Hackenberg et al. \cite{hackenberg2015energy}.
If the communication time of the MPI primitive is longer than 500$\mu$s when the timer callback is triggered the lowest P-state is selected, otherwise the timer is canceled. This timeout-based policy has been in-depth analyzed in the power management literature and has been shown to be effective in mitigating the issues related to prediction inaccuracy and predictive model overfitting \cite{benini_power_survey}. It must be noted that similarly to \fermata, the \cntd{} approach does not distinguish between slack and copy time and execute both of them at the minimum P-state, causing additional overheads.



Hence, in this work, we propose \greta, a novel technique that, while inspired by the methodology proposed in  \cite{cesarini_andare_18,cesarini_countdown}, differently from state-of-the-art approaches induces negligible overhead (in the worst case less than $3\%$) in applications running on real production HPC machines. \greta{} implements purely reactive mechanisms, thus it is robust to miss-prediction errors and capable of isolating slack time with a new reactive approach based on artificial barriers insertion. We will discuss these aspects in Sections \ref{sec:background} and \ref{sec:greta}.

\section{Background}
\label{sec:background}
The proposed manuscript shares some common assumptions with approaches in the state of the art. We list in this Section common definitions, as well as a taxonomy to compare our solution with previous works in this field.

\subsection{Definitions and Assumptions}
\label{sec:def_ass}
We target typical HPC scientific applications, which are usually composed of parallel processes (from tens to thousands) running on a cluster of compute nodes interconnected with a high-bandwidth low-latency network. Each application process is statically bound to a compute element for its entire life duration. Processes can exchange data through the network interconnection using a message-passing interface (MPI) library that can send explicit messages. Multiple processes can share the same compute node since modern HPC machines are equipped with multi- and many-core high-end processors. 
MPI library abstracts the locality of computing resources by taking care of the communication inter and intra nodes. HPC applications can ignore the locality of the processes because MPI  represents the computation resources as a large set of single-core nodes. HPC users can choose the binding configuration of the processes to optimize communication. Non-uniform memory access (NUMA) plays an essential role in terms of communication latency and bandwidth in MPI communication primitives.

In this work, we do not consider heterogeneous compute nodes such as GPU- and MIC-based architectures, leaving them to future extensions of this work. We recall that today (Nov.18 Top500 list \cite{Top500}) $73\%$ (364/500) of worldwide supercomputers systems are homogeneous x86 based clusters.

\begin{figure}
    \centering
    \includegraphics[width=1.00 \linewidth]{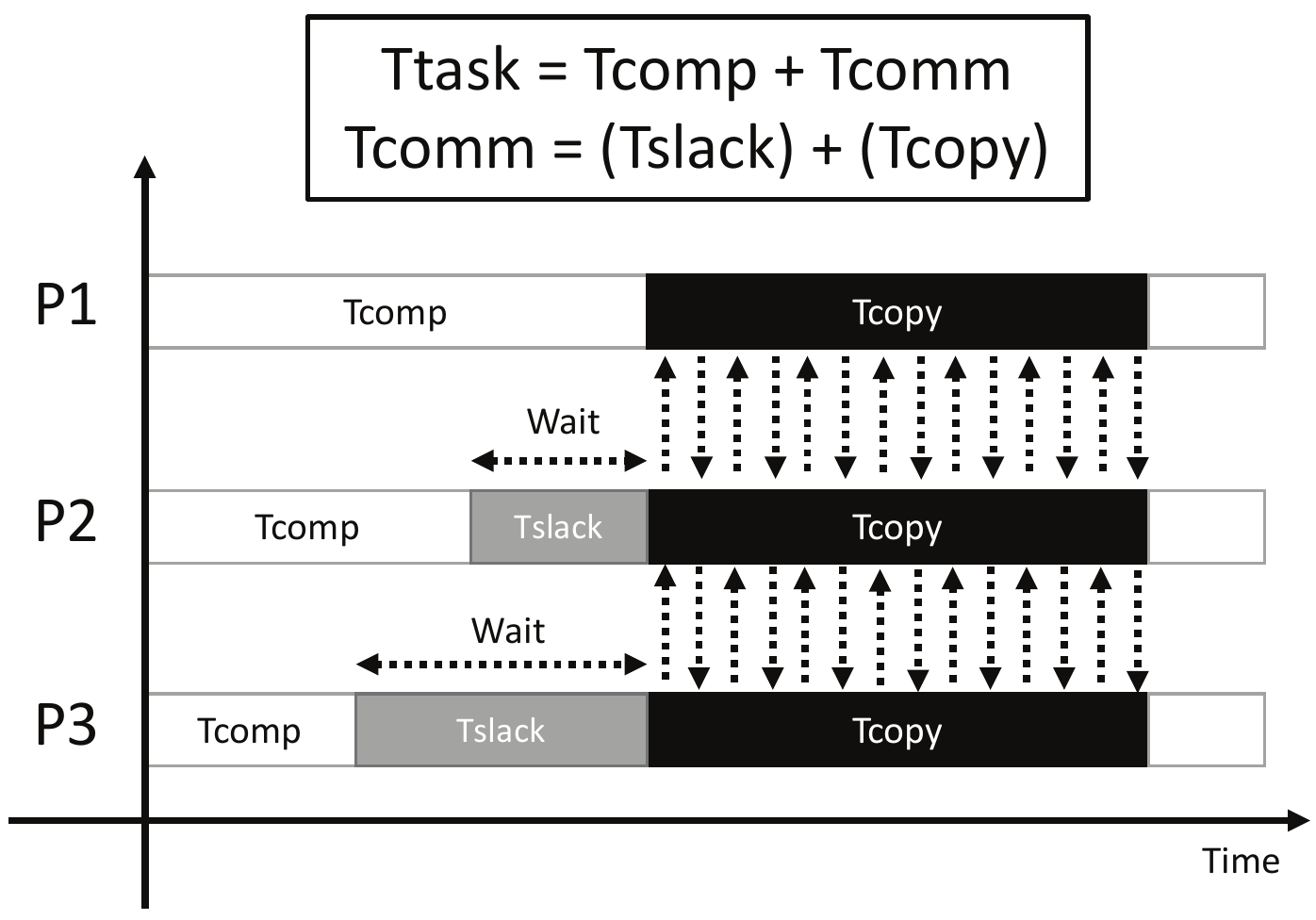}
    \caption{Execution model and taxonomy. A task is shown for each process.}
    \label{fig:exe_model}
\end{figure}

Figure \ref{fig:exe_model} shows our execution model and taxonomy, which is the same of Rountree et Al. \cite{adagio_dynamic}. At the base of the execution model, there is the \textit{task}; each process is composed of a set of tasks that sequentially enter in the execution flow. Each process starts with the MPI primitive \textit{MPI\_Init}, which begins the first task of the process, and ends with the \textit{MPI\_Finalize}, which concludes the last task. Each task comprises (i) a computation time \tcomp, which identifies the time spent in the application code, and (ii) a communication time \tcomm, which is the time spent in the MPI library. In turn, \tcomm{} is composed of (i) a slack time \tslack{} and (ii) a copy time \tcopy. The slack time is the period spent in the MPI library while waiting for the last process that encounters the MPI primitive. This time is purely busy waiting and is equal to zero for the last process. When the last process enters in the primitive, it unlocks all the others. When \tslack{} is concluded, processes can start sending and receiving data -- we call this period copy time. The MPI specification also implements pure synchronization primitives (such as \textit{MPI\_Barrier}, \textit{MPI\_Wait}, etc.) which have zero copy time but cause slack time. In this work, we do not target non-blocking and no-synchronization MPI primitives since they do not produce busy waiting wasted cycles in the application. We define the \textit{critical process} as the last process that enters the MPI primitive; this process plays a critical role as it blocks all the others. A \textit{critical path} is a list of \textit{critical processes} in successive blocking MPI primitives.

\subsection{Power Management Basics}
\label{sec:pm_basic}
Today's generation of high-end CPUs is composed of many processing elements that can work at different voltage and frequency. In particular, Intel technology started to integrate per-core fully-integrated voltage regulators from Haswell architecture \cite{HASWELL}, which allow the DVFS mechanism to trade-off performance and energy. While DVFS control is available in most modern high-end processors for HPC system, our approach explicitly targets Intel architectures for two reasons: i) our target machine is an Intel-based system and ii) most of HPC system in the Top500 list ($73\%$ in Nov 2018 \cite{Top500}) are based on Intel CPUs.

The power control unit (PCU) of Intel architectures is the HW component that controls the power management knobs and exposes model-specific registers (MSR) concerning the DVFS control knobs. While the internal logic of Intel PCUs is not publicly available, Hackenberg et al. \cite{hackenberg2015energy} analyzed the behavior of the DVFS control registers of the PCU in Haswell architectures. Their experimental results show that frequency changes occur at regular intervals of about 500$\mu$s. As pointed out by Cesarini et al. \cite{cesarini_andare_18,cesarini_countdown} this interval creates uncertainties in P-state transitions for code regions shorter than 500$\mu$s. 

It is possible to interact with the DVFS mechanism using the MSRs. 
MSRs are not only used to interact with the HW power manager, but also with the performance counters, debugging, and trace controls. \greta{} requires read and write access to these registers.
Intel provides two specific assembly instructions to read and write MSRs, named respectively \textit{rdmsr} and \textit{wrmsr}. Both instructions are executed in ring 0 (kernel mode), so only the operating system (OS) can execute them. The Linux OS allows user-space access through a kernel driver called \textit{MSR driver}. The drawback of the \textit{MSR driver} is that only the root user can access to this driver because exposing all MSRs to a generic user can lead to security issues. 


The \textit{MSR\_SAFE driver} \cite{MSR_SAFE} overcome the restricted privilege issue and security risks, by supporting a registers white-list. In \greta{} we white-listed a limited subset of control registers in \cite{MSR_SAFE} to let standard HPC users interact with the HW power manager and performance counters.


\subsection{Power Management Modelling}
\label{sec:pm_model}
The majority of HPC power management policies fall into two categories, proactive and reactive policies. Both policies aim at scaling down the P-state in regions of code which are less susceptible (or even not sensitive at all) to frequency scaling. We now report the most common implementation concepts for both categories.

\subsubsection{Proactive Policies}
\label{sec:proactive_pol}
Training strategies are often at the base of proactive policies. A typical train strategy initially starts by identifying the portion of the code that can be targeted for frequency reduction. The runtime always executes a newly encountered code region at the highest available P-state to measure the performance of that region. At the end of the code portion, the runtime stops the performance monitor and gathers the performance parameters; these will be used by the algorithm to compute a new P-state for the next time the code region is encountered. The most frequently used parameters are the execution time of the code region at a given frequency and the number of retired instructions. The next time the execution flow encounters a previously recorded code region, the policy tries to scale the P-state, by applying the earlier computed P-state. A code region can be any part of the execution flow of the program, such as application code \cite{adagio_dynamic} or communication runtime \cite{lim2006adaptive}. The algorithms used to identify the optimal frequency can be a simple last-value prediction \cite{adagio_dynamic} or many complex predictors, like an auto-regressive moving average. A typical implementation of this policy is done through a history table used by the algorithm to predict the next P-state to be assigned. The code regions can be uniquely identified using different strategies: (i)  source code instrumentation \cite{freeh2008just}, (ii) compilers automatic insertion \cite{score_p} or (iii) identified at execution time via a stack trace mechanism \cite{adagio_dynamic,lim2006adaptive}. The advantage of using source code tagging is the low overhead and precise code pinpoint that developers can optimize. However, this methodology requires to modify the application source code which is not always tolerated.
On the contrary, compiler tagging does not require programmer intervention. The drawback is the need to re-compile the source code. Conversely, the stack trace mechanism is completely application-agnostic, and it does not require source code modification nor re-compilation, but extra cycles of computation in the application. As we will see in the experimental results when this is done synchronously to the MPI calls its overhead can be neglected.


\subsubsection{Reactive Policies}
\label{sec:reactive_pol}
Differently from the previous ones, reactive policies are implemented as event-based strategies. When specific events occur, the runtime triggers well-defined actions. Cesarini et al. \cite{cesarini_andare_18,cesarini_countdown} developed \cntd, which implements a reactive policy based on a timeout to filter out code portions which have too short to cause a P-state transition, this depends on the HW PCU delay of Intel architectures. 
Contrary to proactive ones, reactive policies do not need to uniquely identify regions of code, since they do not maintain a history of the execution traces. However, similarly to the proactive ones, the runtime requires to intercept events related to the beginning and end of each region. This can be done by leveraging the profiling extensions of communication libraries such as PMPI or debug symbols \cite{extrae}.

While proactive policies can modify their behavior to adapt to different code regions and minimize overhead, reactive policies always apply the same operation, since they are unaware of the different sensitivities to the frequency scaling of different code portions.
Similarly to \cntd, \greta{} applies a reactive policy to code regions without considering their sensitivity to frequency scaling.

\section{COUNTDOWN Slack - a Low-overhead, Reactive, Slack-Aware PM Runtime}
\label{sec:greta}
In this Section we present the implementation of \greta.

\subsection{Runtime}
\greta{} is a simple shared library written in C language. It can instrument standard MPI-based applications that load \greta{} in their \textit{LD\_PRELOAD} environment variable before the execution of the program. Using this technique, every MPI call is intercepted by \greta{} which executes between the application and the MPI library. It implements a PMPI interface to wrap all the MPI primitives defined in the MPI specification v3. The library has been designed to have the lowest possible overhead and to interact with the hardware through the \textit{MSR\_SAFE} kernel driver as discussed in Section~\ref{sec:pm_basic}. \greta{} also provides a static link version, which can be used when dynamic linking is not possible, to inject \greta{} in the application binary at compilation time. If dynamic linking is allowed, \greta{} does not require any modifications of the source code nor toolchain, nor re-compilation steps. It is completely transparent to the user. In the experimental results of this paper, we instrument all the target HPC benchmarks using dynamic linking.

\greta{} is based on a simple but effective strategy to reduce energy consumption in production HPC systems without performance penalties. The key idea is to scale down the P-state in slack times of the application reducing the frequency but leaving unaltered the performance for both computation and data copy regions. 

Since \greta{} targets performance-neutral energy savings, our goal is to avoid performance penalties for a large set of MPI-based applications, thus \greta{} focuses on saving energy only when this has no effects on performance skipping potential energy saving if they could induce non-negligible performance overhead. As side effect, \greta{} shows slightly lower energy saving respect to the state-of-the-art approaches but guarantees better performance which is the first goal in HPC applications. To accomplish this task, \greta{} employs a purely reactive policy for both slack isolation and short region filtering since predictive algorithms can induce performance penalties in case of mispredictions \cite{kerbyson2011energy}.

\begin{figure*}[t]
    \centering
    \includegraphics[width=1.00 \linewidth]{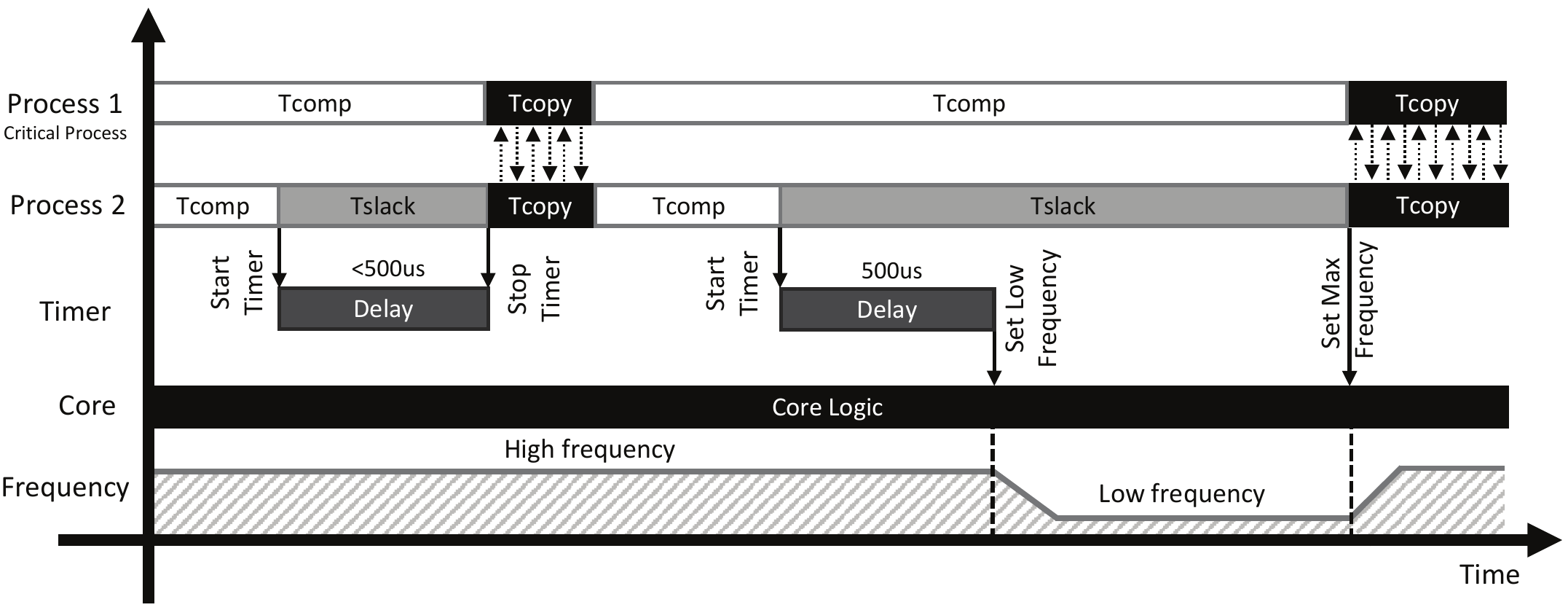}
    \caption{Timer strategy utilized in COUNTDOWN to filter-out short slack times.}
    \label{fig:callback}
\end{figure*}

As shown in \cite{hackenberg2015energy}, modern Intel architectures allow core frequency changes only every 500us. Thus, we implement a timeout policy as presented in \cite{cesarini_andare_18,cesarini_countdown}, but we apply it only to slack times, without varying the cores' frequency during the copy, as shown in Figure \ref{fig:callback}.

\subsection{Reactive Slack Isolation}
The slack time of an MPI call is usually included in the primitives. Differently from previous works that use pre-characterization or non-causal models to separate slack and copy time from the communication time \cite{adagio_static,adagio_dynamic,lim2006adaptive}, in \greta{} we propose a novel reactive approach based on the insertion of artificial/instrumental barriers. This mechanism is agnostic to the MPI implementation, and it is built on top of standard MPI primitives. It can be used with every MPI library. We applied this mechanism on blocking MPI primitives leaving unaltered non-blocking, one-sided, file and support MPI primitives. We also account for collective and P2P (Point-to-Point) primitives since the time spent in other primitives are negligible for the considered benchmarks. We implement two different mechanisms to isolate the slack time, one for collective and one for P2P primitives.

\subsubsection{Collective Barrier}
We designed a straightforward mechanism to separate slack and copy time in collective primitives. Every time the application calls a collective primitive, \greta{} intercepts the call, forces a \textit{MPI\_Barrier} on the same communicator, and reduces the P-state to the minimum available one if the \textit{MPI\_Barrier} is long enough (Reactive Short Phase Filter). When all processes reach the collective primitive, the barrier inserted by \greta{} terminates and the execution flow returns to \greta. After that, \greta{} profiles the slack time, restores the maximum frequency and calls the first collecting primitive of the application. We call this mechanism \textit{Collective COUNTDOWN Slack barrier}.

\subsubsection{Point-to-Point Barrier}
The mechanism for collective barriers is straightforward since all the ranks in the communicator are involved in the barrier. Unfortunately, the P2P primitives are called only from the processes involved in the communication. We cannot insert a \textit{MPI\_Barrier} because it would cause a deadlock on the processes involved in the P2P communication. To overcome this limitation, we implemented a waiting mechanism based on non-blocking primitives.
Before a \textit{MPI\_Send} primitive \greta{} adds an artificial \textit{MPI\_Isend} followed by a \textit{MPI\_Wait}. Similarly before a \textit{MPI\_Recv} primitive \greta{} adds an artificial \textit{MPI\_Irecv} followed by a \textit{MPI\_Wait}. Differently, before a \textit{MPI\_Isend} primitive \greta{} adds only an artificial \textit{MPI\_Isend}, and before a \textit{MPI\_Irecv} primitive \greta{} adds only an artificial \textit{MPI\_Irecv}.

The non-blocking P2P primitive returns back a request object that \greta{} uses in the following \textit{MPI\_Wait} primitive. The \textit{MPI\_Wait} is a blocking primitive used to wait for the completion of a request object. When the application enters in the artificially \textit{MPI\_Wait}, \greta{} reduces the frequency if the \textit{MPI\_Wait} is long enough (Reactive Short Phase Filter).
This mechanism allows \greta{} to obtain a P2P barrier just between the processes involved in the P2P communication and to isolate its slack to copy time. We call this mechanism \textit{P2P COUNTDOWN Slack barrier}. When all processes reach the P2P primitive, the artificial barrier terminates and the execution flow returns to \greta. After that, the library restores the maximum frequency and calls the original P2P primitive of the application.

\greta{} instrument blocking P2P primitives with correspondent non-blocking ones but the application can use mixed blocking and non-blocking P2P primitives and this can create a mismatch of \textit{P2P COUNTDOWN Slack barrier}.
To avoid it, we added non-blocking P2P primitives in-front-of every non-blocking P2P primitives called from the application to balance the number of non-blocking P2P primitives.

To measure the overhead for both \textit{collective and P2P COUNTDOWN Slack barrier}, we run all our benchmarks with and without the barrier mechanism and we compare the execution times. The experimental results show a negligible overhead for all our benchmarks.

\subsection{Reactive Short Phase Filter}
As we showed in \ref{sec:pm_basic}, it is not possible to ensure P-state transitions in code regions shorter than 500$\mu$s. To filter out slack regions shorter than this value, we implemented a timeout policy in \greta{} as the one proposed in \cite{cesarini_andare_18,cesarini_countdown}.
The timeout strategy of \greta{} relies on the standard timer APIs of Linux systems. Linux provides the kernel calls \textit{setitimer()} and \textit{getitimer()} to manipulate Linux timers. The timer allows users to register a callback function; when the timer expires, a system signal interrupts the ``normal'' execution, and the callback is executed. The callback sets the lowest P-state and return. 
At the end of the slack region, \greta{} restores the highest P-state. This mechanism is shown in Fig.~\ref{fig:callback}.

\subsection{Profiler}
\greta{} is endowed with a profiler module that allows a detailed analysis of the application; the profiler is split into two components.

i) The \textit{event profiler} monitors the HW performance counters using the \textit{RDPMC} instruction, reading the performance monitoring units of Intel's processors. \textit{RDPMC} is a low-overhead user-space assembly instruction that can be used to keep track of micro-architectural events at a very high frequency with negligible overhead. This instruction reads the fix performance counters which counts the number of clock cycles at the nominal frequency, at the current P-state, and the instructions retired. Furthermore, it can read a limited subset of configurable HW performance counters used to monitor user-specific micro-architectural metrics. This profiler is also able to extract MPI information from the parameters passed to the MPI primitives.

ii) The \textit{time-based profiler} collects a broad set of HW performance counters every second. This profile is time-based, and it leverages a timer to sample the entire node. Every second, the MPI processes on the same node alternately sample all the core and uncore performance registers in a round-robin fashion. This strategy is used to distribute the profiling overhead among all processes. The profiler maintains the tracing information in a memory area shared among all the processes. The profiler exploits the \textit{MSR\_SAFE} kernel driver to access the performance registers using the batch mode \cite{msr_safe_batch} to reduce the overhead. Moreover, it uses Intel Running Average Power Limit (RAPL) registers to monitor the energy/power consumed by the CPU and DRAM. The energy measurements presented in the rest of this work always refer to both package and DRAM consumption.

The profiler does not save all the events and time-based traces, but it summarizes them in a hierarchical report. This report comprises a summary file with information about the entire application run, an MPI report with information about MPI primitives, and a set of reports organized for nodes, sockets, and cores. These reports contain the same information of the summary report (plus specific metrics) but organized, respectively, for nodes, sockets, and cores. The hierarchical organization improves the readability of the reports and their long term compression. The overhead of the hierarchical report is entirely negligible, while the overhead of the event and time-based profilers are strictly related to the performance of the storage. In our target architecture with a small set of computing nodes (29), both event and time-based tracing overhead are negligible. The memory footprint of the profiler is constant due to the fixed number of performance counters. It is in the order of few megabytes per MPI process. We recall that the different profiler modalities can be easily configured and deactivated.

\section{State-of-the-art Energy-aware Runtimes}
\label{sec:energy_runtime}
In this Section, we discuss the energy-aware runtimes that we have implemented as part of \greta{} for comparisons with the state of the art. In Section \ref{sec:exp_results}, we compare \greta{} with the following described algorithms.

\subsection{Fermata}
The first runtime we introduce for comparison with \greta{} is \fermata{} \cite{fermata,adagio_dynamic}. \fermata{} implements a simple algorithm to reduce the cores' P-state in communication regions (\tcomm). \fermata{} uses a prediction algorithm to decide when scaling down the P-state; the prediction is determined by the amount of time spent in communication during the previous call. If the duration is greater than or equal to twice the switching threshold, \fermata{} sets a timeout to expire at the threshold time. The threshold time is empirically set to 100ms. Calls are identified as specific MPI primitives in the application code through the hash of the pointer that makes up the stack trace. The hash is generated when the application encounters an MPI primitive; hence, each MPI primitive in the code is uniquely identified. The information about the last call is stored in a look-up table used to choose if to set the timer in the next call.

In \greta{} we implemented two versions of the Fermata policy, one with the original empirical switching threshold value of 100ms \cite{adagio_dynamic}, and one with an empirical switching threshold tuned for the target system of 500$\mu$s \cite{hackenberg2015energy}.

\subsection{Andante}
Differently from \fermata{}, \andante{} \cite{adagio_dynamic} focuses on slowing down the computation region (\tcomp) to reduce the slack time (\tslack). This approach is based on the assumption that \tcomp{}, \tslack, and the number of instructions retired for a given task will be the same as the previous one for the same task. \andante{} computes the highest P-state for \tslack{} (aiming at reducing it) by exploiting the instructions per second (IPS) estimated based on measurements the previous time the same task was encountered (last-value prediction).
\andante{} logic in \cite{adagio_dynamic} uses a pre-characterization of the message-transfer time of the MPI library to estimate the \tcopy. \tslack{} is calculated as the difference between \tcomm{} and \tcopy. Similarly to \fermata, \andante{} distinguishes tasks using the stack trace at the end of each collective MPI primitive. The information regarding the last executed task is kept in a look-up table containing the IPS for each discrete P-state of the system and the next P-state to assign.

\greta{} implements the same logic of \andante, but is uses the \textit{Collective and P2P COUNTDOWN Slack barrier} to compute \tslack{} as the \cite{adagio_dynamic} pre-characterization step cannot be ported as it is in NUMA compute node. 

\subsection{Adagio}
The idea behind \adagio{} \cite{adagio_dynamic} is to merge \fermata{} and \andante{} in a single energy-aware runtime. \andante{} slows down the computation regions, while \fermata{} handles the communication phases. 

In \greta{} we implemented the same logic of \adagio{} by combining \andante{} and \fermata{}. We used in this case only \fermata{} configured with the empirical switching threshold at 500$\mu$s and applied only to the slack regions isolated with the \textit{Collective and P2P Countdown Slack barrier} logic.

\subsection{COUNTDOWN}
\cntd{} \cite{cesarini_countdown} is a runtime library to identify and automatically reduce the power consumption of the computing elements during the communication phases. 
It uses a timeout strategy to filter-out short communication regions (those too fast for the DFVS control knob to react, i.e., shorter than 500$\mu$s). \cntd{} differs from \greta{} as it considers the 
communication phase, while \greta{} focuses only on the slack time. However, the timeout implementation is similar.

\section{Experimental Results}
\label{sec:exp_results}

\subsection{Experimental Setup}
\label{sec:exp_setup}

\subsubsection{Target Architecture}
For all the experiments we use a Tier-1 HPC system based on an IBM NeXtScale cluster which is currently ranked in the Top500 supercomputer list \cite{Top500}.
The compute nodes of the HPC system, are equipped with 2 Intel Broadwell E5-2697 v4 CPUs, with 18 cores at 2.3 GHz nominal clock frequency and 145W TDP and 128 GB of DDR4. Each node runs the Centos 7 OS and Linux kernel 3.10.0, nodes are interconnected with an Intel QDR (40Gb/s) Infiniband high-performance network.
We compile all our benchmarks using \textit{GCC/GFortran 6.2} as our toolchain, coupled with \textit{OpenMPI 3.2} as the communication library.

The default configuration for the power management in the target system is with the Linux \textit{cpufreq} driver at the maximum P-state with turbo mode enabled. This is the baseline for our experimental results and we refer to this configuration lately as \plain. 

\begin{table}
    \small
    \sf
    \centering
    \resizebox{\columnwidth}{!}{
        \begin{adjustbox}{width=1\textwidth}
        \begin{tabular}{l rrr rrr}
        \toprule
        \multirow{2}{*}{\emph{Application}} & \multicolumn{3}{c}{\emph{Without}
        Previous Info} & \multicolumn{3}{c}{\emph{With} Previous Info} \\ 
        & \tcomp & \tslack & \tcopy & \tcomp & \tslack & \tcopy \\ 
        \midrule
        nas\_bt.E.1024 & 57.0 & 17.6 & 52.5 & 6.2 & 12.4 & 12.4 \\
        nas\_cg.E.1024 & 21.9 & 7.1 & 25.3 & 16.2 & 5.5 & 11.0 \\
        nas\_ep.E.128 & 9.1 & 8.4 & 23.8 & 9.7 & 7.3 & 24.6 \\
        nas\_ft.E.1024 & 1.2 & 5.4 & 9.7 & 0.3 & 1.2 & 3.9 \\
        nas\_is.D.128 &  10.7 & 15.2 & 8.2 & 5.3 & 8.0 & 2.4  \\
        nas\_lu.E.1024 & 0.9 & 19.8 & 0.5 & 0.7 & 13.5 & 0.4 \\
        nas\_mg.E.128 & 5.1 & 4.8 & 13.0 & 4.1 & 5.3 & 13.1 \\
        nas\_sp.E.1024 & 46.5 & 11.8 & 46.9 & 4.1 & 10.2 & 7.3 \\
        omen\_1056p & 1.0 & 57.3 & 75.8 & 2.8 & 55.4 & 64.6 \\
        \midrule
        \emph{Average} & 17.0 & 16.4 & 28.4 & 5.5 & 13.2 & 15.5 \\  
        \bottomrule
        \end{tabular}
        \end{adjustbox}
    }
    \caption{Prediction error [\%] for all test applications (SMAPE)}
    \label{tab:pred_errors_smape}
\end{table}

\subsubsection{NAS Parallel Benchmark}
The NAS Parallel Benchmark suite (NPB) is a set of popular HPC benchmarks developed by the NASA Advanced Supercomputing division. The NPB consist of 8 benchmarks and kernel namely BT, CG, FT, LU, SP, EP, MG and IS, which are widely used in different scientific areas such as spectral transform, fast Fourier transform,  partial differential equations, fluid dynamics, and so on. We used NPB version 3.3.1 and tested different configurations to balance the duration of all benchmarks at around 10 minutes of execution time. For BT, CG, FT, LU, and SP we ran on 29 nodes using 1024 cores with data set E while for EP and MG we used the same dataset but on four nodes and 128 cores. Instead, for IS we use dataset D, which is the largest available one for that benchmark.

\subsubsection{OMEN}

OMEN is an atomistic quantum transport simulator that can compute the I-V characteristics of all kinds of nano-devices at the ab initio level (from first principles) \cite{Luisier_1,Luisier_2}. The code has been optimized to run on the largest available supercomputers, reaching two times the ACM Gordon Bell Prize final \cite{Luisier_3,Luisier_4}. Here, a transistor with a 2-D crystal as channel material serves as a benchmark. 

We tested two configurations for OMEN, the above-mentioned called OMEN.1056p, which it runs on 29 nodes using 1056 cores and the OMEN.60p which runs on a couple of nodes with a scale-down dataset.


\subsection{Regions Predictability}
\label{sec:phases_predictability}

In this subsection, we report an analysis carried out on the test applications to highlight the degree of predictability of \tcomp, \tslack, and \tcopy, based on features available before a given region is encountered. As depicted in Section \ref{sec:background} most of the state-of-the-art energy-management runtimes use a predictor to estimate the duration of the regions of an application for optimization purposes. To assess the predictability of the region duration, we employ a standard Machine Learning (ML) technique, namely Random Forest (RF) \cite{Breiman:2001:RF:570181.570182} models\footnote{Random Forests are ensembles of decision trees}. For each test application, we build and train three RF models, one for each target region duration (\tcomp, \tslack, and \tcopy); we then evaluate the quality of the prediction by measuring the difference between the real region duration and the estimates (over a test set of examples not seen during the training phase). All the experiments on the predictability were performed using the \emph{scikit-learn}\cite{scikit-learn}, a widespread ML library for python.

For this purpose we extracted with \greta{} \textit{Event Profiler} a set of traces for the NPB and OMEN benchmarks in the default configuration (\plain{}). For each benchmark (8 NPB + OMEN) we obtain a trace with a row for each code region and a column for each feature analyzed. The rows are ordered first on the rank id, then on the task id, and finally on the progression of time. 

\begin{figure}
    \centering
    \includegraphics[width=.48\textwidth]{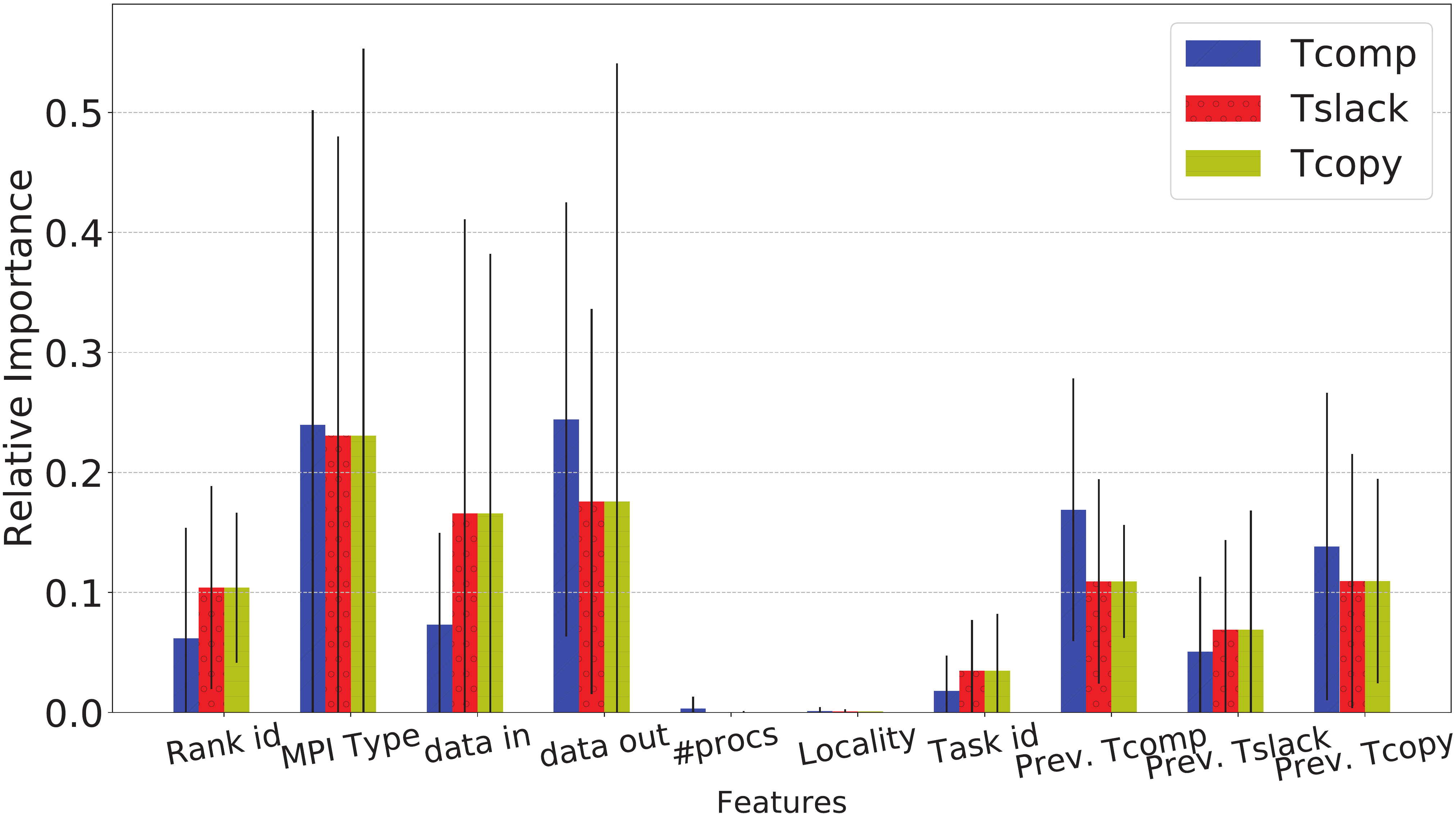}   
    \caption{Relative importance of each feature (including information about
    previous phases); the colors identify the prediction targets}
    \label{fig:feat_importances_all}
\end{figure}

We consider two types of approach, namely a first one taking into consideration only the features relative to the MPI call whose region duration we want to predict, and a second one that exploits also information about the last MPI call of the same type and rank. 
We now describe the first approach. Each data set is composed of the following features: 
1) rank id of the process that makes the MPI call; 
2) type of the MPI call; 
3) size (byte) of the received data; 
4) size of the sent data (byte); 
5) number of processes involved in the MPI call; 
6) locality, a number between 0 and 1 that specifies the amount of
local (to the node) or remote processes involved in the call (0 means that all
processes are remote, 1 all local processes); 
7) task id, the hash of the stack (identify MPI calls made at the same points of code). 
Each RF model aims at predicting one of the three targets, \tcomp, \tslack, and \tcopy{} (expressed in microseconds). We limit our analysis to processes with a duration longer than 500ms, we are less interested in extremely short MPI calls since they offer lower potential in terms of power management. There is a distinct data set for each test application, with sizes of varying dimensions, ranging from 1k elements to 1 million examples. We recall that the first approach is motivated by the state-of-the-art works and uses pre-characterization of the copy time to estimate at execution time the slack time. Indeed the  \tcopy{} model can be seen as a generalization of the pre-characterization steps.

The second approach is motivated by state-of-the-art methods that exploit information regarding previous MPI calls to estimate the phase duration (last value prediction). More precisely, in this case, each example in the data set possesses the set of aforementioned features plus three more fields: the \tcomp, \tslack, and \tcopy{} values of the last MPI call with the same type, task, and rank. In both approaches (with or without previous MPI call information), each data set is split in a training set (70\% of the whole set) used to train the RF models, and a test set used to evaluate the quality of the predictions. After a preliminary analysis, we discovered that the RF model accuracy increased if we used the natural logarithm of the target during the training phase, rather than directly using the duration in microseconds; probably, this happens because the logarithm flattens the peaks caused by extremely long or short duration (with respect to the average phase duration for the training set). The accuracy results reported later (test phase) are instead computed on the actual values, by applying the exponential function to the predictions made by the RF models.

In Table~\ref{tab:pred_errors_smape} we see the results of the predictability experiments. The prediction error is computed as percentage, in particular using the SMAPE metric (Symmetric Mean Absolute Percentage Error), preferred to the standard mean absolute percentage error since the former places smaller emphasis on regions with shorter duration\footnote{For a single prediction it is computed as: $SMAPE = 100 \cdot \frac{abs(pred -actual)}{pred + actual}$, where $pred$ is the predicted region duration and $actual$ is the real value. If we use only the actual value at the numerator, as in the standard mean average percentage error, examples with very short duration would significantly skew the overall error}. Each row corresponds to a different test application (identified by the first column); the final one reports the average over all applications. Columns 2-4 show the error obtained if we train ML model without providing information about the duration of previous regions; columns 5-7 report the results obtained
\emph{with} previous region information. The pairs of three columns \tcomp, \tslack, and \tcopy{} correspond to the three region-targets -- one triplet for each approach (with/without previous information).

Table~\ref{tab:pred_errors_smape} reveals that it is not straightforward to predict the 
target phase duration with the collected information. If we consider the 
case without previous MPI call information and we look at the average values 
computed over all applications, the error for predicting both
 \tcomp{} and \tslack{} is around 16-17\%, while for \tcopy{} the accuracy is 
 even lower (28\% error). As one could have expected, things improve if we 
 feed the ML models with additional information regarding the last MPI calls 
 of the same rank, task, and type; this is especially true for \tcomp{} (error 
 decreased at around 5\%) and \tcopy, while the improvement for \tslack{} is 
 less significant. Generally speaking, considering previous information 
 improves the prediction accuracy on most test applications (even drastically:
 see for example \emph{nas\_bt.E.1024} and \emph{nas\_sp.E.1024}).
However, this is not true for all applications and in some cases (for instance
 \emph{omen\_1056p}) the additional information can lead to a marginal 
 decrease in prediction accuracy, probably due to an increase in noise that 
 confuses the RF models (however, the effect is marginal).
 
Another aspect that deserves more analysis is understanding the factors that the RF models focus on in order to make their estimates. We can gain some insight into this matter by inspecting the importance of each feature. 
In \emph{scikit-learn} the feature importance is computed as mean decrease 
in impurity, described in \cite{breiman1984classification}; this method is known 
to be prone to bias (see\cite{strobl2007bias}), thus we opted to compute the 
feature importance via a permutation-based approach 
\cite{strobl2008conditional,parr2018beware}, 
where each feature importance is computed by looking at how random 
re-shuffling (which preserves the feature distribution) influences the model 
accuracy. We normalized the important values in the [0,1] range; zero 
indicates that a feature has no importance for the regression RF model, while 
 values closer to 1 indicate relatively more important features.

Figure~\ref{fig:feat_importances_all} shows the importance of all features when we include information about previous MPI calls. The plot refers to the average feature importance computed overall test applications. The height of each bar corresponds to the average feature importance and the black vertical line represents the standard deviation -- longer black lines indicate that the feature importance varies significantly for different applications. Each target
 is highlighted by a different color, blue for \tcomp, red for \tslack, green for
 \tcopy. Each group of three bars represent a feature; the three right-most groups report the influence of the previous MPI call (same type) regions duration, respectively (from left to right), last MPI call \tcomp, \tslack, and \tcopy.
 
 The first thing that we can notice is that the standard deviations tend to be quite large, revealing a high variability. This high variability suggests that predicting the region duration of MPI calls in HPC application is indeed a difficult task since there is no subset of trustworthy features that can be robustly employed for accurate estimation.
 Secondly, the number of processes, the locality, and the task id have very little influence on the RF prediction. We recall, that the task id was one of the foremost important feature in state-of-the-art approaches \cite{lim2006adaptive,adagio_dynamic}.
 
 To summarize, the features with greater importance are the size of the 
 outgoing transmitted data (especially for the \tcomp) and the type of the MPI call.
 The size of the incoming data is more relevant for \tslack{} and 
 \tcopy. Additionally, the importance of the information about the previous MPI call with the same type cannot be discounted, as highlighted by the right-most three groups of columns. In particular, the length of \tcomp{} and \tcopy{} phases of the last MPI call is an important factor for the models that predict \tcomp.  

As we mentioned in Section \ref{sec:greta}, \greta{} employs only reactive mechanisms to isolate slack and filter out too short slack regions to avoid miss-prediction errors.  





\begin{table}
    \small
    \sf
    \centering
    \resizebox{\columnwidth}{!}{
        \begin{adjustbox}{width=1\textwidth}
        \begin{tabular}{l|rrrrrrr}
        \toprule
        Application & Tcomm & Tslack & \makecell{\fermata\\100ms} & \makecell{\fermata\\500$\mu$s} & CNTD & \makecell{CNTD\\Slack} & \makecell{AVG MPI\\Time Duration} \\ 
        \midrule
        nas\_bt.E.1024 & 0.12  & 0.07  & 0.00  & 0.00  & 0.12  & 0.07  & 1.831 \\
        nas\_cg.E.1024 & 34.84 & 0.07  & 0.39  & 32.68 & 32.96 & 0.01  & 2.068 \\
        nas\_ep.E.128  & 7.56  & 7.56  & 0.00  & 0.00  & 7.56  & 7.56  & 24384.882 \\
        nas\_ft.E.1024 & 65.10 & 12.28 & 55.88 & 57.80 & 65.09 & 12.28 & 2374.646 \\
        nas\_is.D.128  & 62.73 & 27.42 & 31.14 & 40.98 & 62.65 & 27.41 & 277.003 \\
        nas\_lu.E.1024 & 51.01 & 45.51 & 9.91  & 21.93 & 22.42 & 21.79 & 0.099 \\
        nas\_mg.E.128  & 8.94  & 0.09  & 0.01  & 7.95  & 8.48  & 0.06  & 1.134 \\
        nas\_sp.E.1024 & 0.05  & 0.02  & 0.00  & 0.00  & 0.05  & 0.02  & 1.447 \\
        omen\_60p      & 59.69 & 56.00 & 43.87 & 48.86 & 59.60 & 55.99 & 59.853 \\
        omen\_1056p    & 62.96 & 56.42 & 50.85 & 60.18 & 62.83 & 56.41 & 58.193 \\
        \bottomrule
        \end{tabular}
        \end{adjustbox}
    }
    \caption{Slack Isolation Potential [\%] and average MPI time duration [ms].}
    \label{tab:slack_isolation}
\end{table}


\subsection{Slack Isolation Policy Performance}
\label{sec:PM_characterization}

As discussed in the previous subsection, it is difficult to obtain accurate predictions for \tcomp, \tslack, \tcopy, and \tcomm. Moreover, the energy savings and performance neutrality depend on the capability of the given algorithm to (i) isolate the slack time from the copy time in MPI primitives, and (ii) avoid P-state transitions for a time shorter than 500$\mu$s\cite{hackenberg2015energy,cesarini_countdown}.

In this subsection, we analyze the capability of \greta{} in comparisons with state-of-the-art approaches in taking advantage of slack time to reduce the energy consumption and limiting the application overhead while discarding shorter slack regions and copy time regions. We conducted this analysis on the application traces recorded by \textit{\greta{} Event Profiler} during the execution of the test applications with default node power management settings (\plain{} configuration). These traces are the same as used in Subsection \ref{sec:phases_predictability}. On these traces we have implemented the \fermata{} and \cntd, as described in Section \ref{sec:energy_runtime}, as well as the \greta{} isolation and timeout policy. For the \fermata{} algorithm we report both versions with the empirical switching threshold set at 100ms (as described in \cite{adagio_dynamic}) and at 500$\mu$s (adapted to the characteristics of the target architecture \cite{hackenberg2015energy,cesarini_countdown}).

Table~\ref{tab:slack_isolation} show the results of this test. Each row corresponds to a different application, and the column reports the total \tcomm{} and \tslack{} time, as well as for each power management runtime analyzed the total time in which the algorithm is capable of reducing the power consumption.
Values in each column are reported in percentage with respect to the execution time of the application. The column \textit{AVG MPI Time Duration} reports the average time duration of the MPI primitives in milliseconds.

\begin{table*}
    \small
    \sf
    \centering
    \begin{adjustbox}{width=1\textwidth}
    \begin{tabular}{l|rrrrrr|rrrrrr|rrrrrr}
    \toprule
    \multirow{2}{*}{\makecell{\\\emph{Application}}} & \multicolumn{6}{c}{\emph{Ex.Time Overhead}} & \multicolumn{6}{c}{\emph{Energy Saving}} & \multicolumn{6}{c}{\emph{Power Saving}} \\ 
    & Min Freq & Fermata & Andante & Adagio & CNTD & \makecell{CNTD\\Slack} & Min Freq & Fermata & Andante & Adagio & CNTD & \makecell{CNTD\\Slack} & Min Freq & Fermata & Andante & Adagio & CNTD & \makecell{CNTD\\Slack} \\
    \midrule
    nas\_bt.E.1024
    & {\color{red}\textbf{72.18}} & 1.95 & {\color{red}\textbf{77.72}} & {\color{red}\textbf{68.94}} & {\color{red}\textbf{8.92}} & 0.75
    & 3.39 & 2.07 & 0.11 & 3.35 & 5.96 & 7.97
    & 43.89 & 3.95 & 43.79 & 42.79 & 13.66 & 8.65 \\
    nas\_cg.E.1024 
    & {\color{red}\textbf{21.73}} & 3.86 & {\color{red}\textbf{8.18}} & {\color{red}\textbf{14.35}} & 4.23 & 1.08
    & 21.59 & 18.89 & 24.72 & 22.69 & 22.58 & 9.57
    & 35.59 & 21.91 & 30.41 & 32.39 & 25.72 & 10.54 \\
    nas\_ep.E.128 
    & {\color{red}\textbf{136.04}} & -0.31 & -0.15 & 1.30 & 0.80 & -0.60
    & {\color{red}\textbf{-15.00}} & 0.62 & 0.10 & {\color{red}\textbf{-1.35}} & 0.05 & 1.04
    & 51.28 & 0.31 & -0.05 & -0.05 & 0.84 & 0.44 \\
    nas\_ft.E.1024 
    & {\color{red}\textbf{34.54}} & 2.57 & {\color{red}\textbf{24.32}} & {\color{red}\textbf{30.22}} & 3.50 & 0.26
    & 20.89 & 23.59 & 18.25 & 17.76 & 25.92 & 6.25
    & 41.20 & 25.51 & 34.24 & 36.85 & 28.42 & 6.50 \\
    nas\_is.D.128
    & {\color{red}\textbf{29.95}} & 3.13 & 3.86 & 4.23 & 3.21 & 1.85
    & 19.42 & 17.89 & 17.63 & 17.82 & 22.65 & 11.32
    & 37.99 & 20.38 & 20.70 & 21.16 & 25.05 & 12.93 \\
    nas\_lu.E.1024 
    & {\color{red}\textbf{77.56}} & {\color{red}\textbf{12.79}} & {\color{red}\textbf{115.86}} & {\color{red}\textbf{144.75}} & {\color{red}\textbf{7.65}} & 3.02
    & 3.82 & {\color{red}\textbf{-9.96}} & {\color{red}\textbf{-15.62}} & {\color{red}\textbf{-24.69}} & 4.30 & 4.16
    & 45.83 & 2.51 & 46.44 & 49.05 & 11.10 & 6.97 \\
    nas\_mg.E.128 
    & 4.15 & 0.52 & 4.09 & 4.29 & -0.14 & 0.03
    & 22.58 & 6.41 & 7.83 & 13.71 & 10.68 & 1.57
    & 25.82 & 7.09 & 11.64 & 17.43 & 10.74 & 1.81 \\
    nas\_sp.E.1024 
    & {\color{red}\textbf{12.44}} & -0.07 & {\color{red}\textbf{5.41}} & {\color{red}\textbf{5.16}} & -0.01 & 0.34
    & 22.28 & 15.12 & 23.71 & 24.11 & 18.62 & 18.44
    & 30.88 & 15.06 & 27.62 & 27.83 & 18.61 & 18.72 \\
    omen\_60p
    & {\color{red}\textbf{120.65}} & {\color{red}\textbf{5.01}} & {\color{red}\textbf{108.65}} & {\color{red}\textbf{114.44}} & {\color{red}\textbf{8.81}} & 0.77
    & {\color{red}\textbf{-9.72}} & 15.12 & {\color{red}\textbf{-20.19}} & {\color{red}\textbf{-14.59}} & 17.33 & 17.14
    & 50.27 & 19.18 & 42.40 & 46.56 & 24.03 & 17.77 \\
    omen\_1056p
    & {\color{red}\textbf{42.12}}	& 2.45 & {\color{red}\textbf{38.59}} & {\color{red}\textbf{41.04}} & 3.22 & 0.38 
    & {\color{red}\textbf{-3.67}}	& 20.99 & {\color{red}\textbf{-2.09}} & {\color{red}\textbf{-4.26}} & 24.72 & 22.11
    & 0.71 & 26.63 & 0.99 & 1.33 & 34.28 & 22.92 \\
    \midrule
    AVG 
    & {\color{red}\textbf{55.14}} & 3.19 & {\color{red}\textbf{38.65}} & {\color{red}\textbf{42.87}} & 4.02 & 0.79
    & 8.56 & 11.07 & 5.45 & 5.46 & 15.28 & 9.96
    & 36.35 & 14.25 & 25.82 & 27.53 & 19.24 & 10.73 \\
    \midrule
    WORST 
    & {\color{red}\textbf{136.04}} & {\color{red}\textbf{12.79}} & {\color{red}\textbf{115.86}} & {\color{red}\textbf{144.75}} & {\color{red}\textbf{8.92}} & 3.02 
    & {\color{red}\textbf{-15.00}} & {\color{red}\textbf{-9.96}} & {\color{red}\textbf{-20.19}} & {\color{red}\textbf{-24.69}} & 0.05 & 1.04
    & 0.71 & 0.31 & -0.05 & -0.05 & 0.84 & 0.44 \\
    \bottomrule
    \end{tabular}
    \end{adjustbox}
    \caption{Comparison of execution overhead, energy, and power saving using different approaches [\%]. We highlighted in bold and red Ex.Time Overhead not negligible ($>5\%$) and energy losses.}
    \label{tab:benchmark}
\end{table*}

From Table~\ref{tab:slack_isolation}, we first notice that for the different applications (rows) the \tslack{} time is a sub set of the \tcomm{} time. For some applications (BT, GC, MG, SP, and FT of the NPB) the \tslack{} time is small fraction of the \tcomm{} timed, while for others (IS, LU and the OMEN benchmark) the \tslack{} time is significant.

As expected, we note that \fermata{} 500$\mu$s outperforms \fermata{} 100ms for all benchmarks as the 100ms empirical switching threshold was extracted by the authors of \cite{adagio_dynamic} on an older system with different power management characteristics than the one used in this study. We also highlight that in moving from 100ms to 500$\mu$s the potential energy saving increases drastically for the NPB, but less for the OMEN production runs. 
When comparing \fermata{} 500$\mu$s with \cntd{} we observe that the reactive timeout policy of \cntd{} is always more effective than the proactive timeout policy of \fermata, leading to remarkable additional energy savings up to $11\%$ more for OMEN.60p and $22\%$ more for the is.D.128 case. It must be emphasized that both \fermata{} and \cntd{} are slack agnostic and thus cannot prevent the policy to slow down also \tcopy{} regions. Differently, the slack isolation policy proposed in \greta{} can separate the \tslack{} regions from the \tcopy{} ones. This is visible in Table \ref{tab:slack_isolation} as \greta{} obtains in general lower coverage of the \tcomm{} and focuses only in the \tslack. We also note that for real application production run as OMEN.60p \greta{} is capable of capturing more power saving opportunities than \fermata{} 500$\mu$s even if \greta{} targets the slack time only. It is also interesting to underline that lu.E.1024, differently from the other applications is characterized by a large fraction of \tcomm{} ($>50\%$), which is almost entirely \tslack{} ($45\%$), but the application spends half of this time in \tcomp{} regions which are shorter than 500$\mu$s (visible from the column \textit{AVG MPI Time Duration}). This can be seen by \fermata, \cntd, and \greta{} which cannot exploit all the \tcomm{} time. 


The next subsection quantifies the energy saving and the overhead mitigation of \greta{} with respect to state-of-the-art approaches.
 



\subsection{COUNTDOWN Slack Run-time Results}
\label{sec:PM_results}
In this subsection we report the performance penalty (if any), power and energy saving of the proposed \greta{} power management runtime with respect to state-of-the-art approaches presented in Section \ref{sec:energy_runtime} when applied to the different benchmarks.

We use as a baseline of our characterization the \plain{} case, we also take into exam the case where all the nodes were configured to operate statically at the minimum available P-state (\mfreq). This is an important scenario as it allows to put in perspective the impact of policies that change the P-state in the computation regions, like \adagio{} and \andante{}. We report for each configuration the execution time overhead (Ex.Time Overhead), the power saving, and the energy saving with respect to the \plain{} case.



By looking at the \mfreq{} we notice that almost all benchmarks (excluded nas\_ep.E.128, OMEN\_60p, and OMEN\_1056p) are memory bound as executing them at the minimum P-state always leads to an energy saving. Moreover, the \mfreq{} case shows, as expected, the highest power saving. This is not true for the energy saving as \mfreq{} causes, in general, a non-negligible overhead in the execution time.

If we first focus on the row of the table named average, which shows the average results for all the benchmarks, we can make the following observations:

(1) \mfreq{} induces the highest overheads in the application execution time. This is expected for \mfreq{} because the entire application is executed at the minimum P-state available.

(2) \andante{} and \adagio{} algorithms reduce the frequency on the \tcomp{} regions of the application, achieve the maximum power saving but lead to significant performance penalty, respectively in average $38.65\%$ and $42.87\%$. This shows that the predictive logic of \andante{} is not capable of effectively estimating in advance the slack of the application regions and their instruction per second in today's real production HPC systems.

(3) The two approaches \fermata{} and \cntd, which are not aware of the \tslack{} but use a timeout based policy, have a performance overhead of $3.19\%$ and $4.02\%$ respectively. While these two approaches have similar time to completion, the energy and power saving is lower on average and \cntd{} achieves an additional $4.21\%$ of energy saving. 
Instead, \greta{} achieves a negligible performance overhead ($<1\%$) with respect to \fermata{} and \cntd{} with a significant energy saving of $9.96\%$.

If we focus on the worst-case results, we can observe:

(1) \mfreq{} does not induce the highest overheads in the application execution time. This because \andante{} and \adagio{} algorithms can induce a non-negligible overhead caused by the hash of the call stack in very short MPI communications (as instance in nas\_lu.E.1024).

(2) The overhead of \fermata{} and \cntd{} approaches can be very significant, respectively $12.79\%$ and $8.92\%$. It must be noted that \fermata{} is worse than \cntd{} for the time needed for computing the hash of the stack used by its prediction algorithm. As effect of this \fermata{} induces an energy penalty of $9.96\%$. While \greta{} is always able to maintain a tolerable overhead for HPC applications, even in the worst case ($\leq3\%$).

(3) \greta{} never induces energy penalty, while all the other approaches induce between $9.96\%$ and $24.69\%$ of energy penalty except \cntd{}.

If we now look at the individual benchmarks, we observe particular features that better describe the benefits of the proposed \greta{} algorithm with respect to the state-of-the-art approaches and \plain{} case.



From the nas\_lu.E.1024, we can see that the \andante{} algorithm induces a severe slowdown, which is even worse than the \mfreq{} case, 
this counter-intuitive result is originated by the overhead related to the task prediction algorithms, which requires to compute the hash of the call stack. This becomes critical in applications with a high density of MPI calls as shown in Table~\ref{tab:slack_isolation}.

We conclude, as also suggested by the previous analysis of the predictability of the region duration, that proactive approaches are not suitable for performance-neutral energy-saving scenarios of supercomputers. On the contrary, the proposed algorithm \greta{} is effective in isolating the slack and filtering out short \tcomp{} regions leading to significant energy saving (up to $22.11\%$ for large-scale production runs) with negligible overhead (always below $3\%$). This proves the effectiveness of the slack insertion logic combined with the timeout policy, making \greta performance-neutral.

\section{Conclusion}
\label{sec:conclusion}

In this paper, we present \greta, a new power management runtime for scientific computing systems. \greta{} combines a novel artificial slack insertion logic with a timeout policy for performance-neutral energy reduction in MPI-based applications. We tested \greta{} in a large set of HPC benchmarks extracted from the NAS parallel benchmark suite and with production runs of the two-times ACM Gordon Bell finalist, OMEN, a quantum-transport application. We compared the proposed approach with reactive and proactive power management libraries presented in the state of the art, showing that \greta{} can preserve the application execution time even in worst cases while reducing the energy consumed by the compute units on average by $9.96\%$. \greta{} allows discovering the communication slacks automatically, reducing the core's frequency, and saving energy. From our findings \greta{} is the only runtime that at the same always leads to an energy saving (proportional to the communication slacks) with negligible execution time overheads ($<3\%$).

\ifCLASSOPTIONcompsoc
  \section*{Acknowledgments}
\else
  \section*{Acknowledgment}
\fi
Work supported by the EU FETHPC project ANTAREX (g.a. 671623), EU project ExaNoDe (g.a. 671578), and CINECA research grant on Energy-Efficient HPC systems.

\bibliographystyle{IEEEtran}
\bibliography{main}

\begin{thebibliography}{10}
\providecommand{\url}[1]{#1}
\csname url@samestyle\endcsname
\providecommand{\newblock}{\relax}
\providecommand{\bibinfo}[2]{#2}
\providecommand{\BIBentrySTDinterwordspacing}{\spaceskip=0pt\relax}
\providecommand{\BIBentryALTinterwordstretchfactor}{4}
\providecommand{\BIBentryALTinterwordspacing}{\spaceskip=\fontdimen2\font plus
\BIBentryALTinterwordstretchfactor\fontdimen3\font minus
  \fontdimen4\font\relax}
\providecommand{\BIBforeignlanguage}[2]{{%
\expandafter\ifx\csname l@#1\endcsname\relax
\typeout{** WARNING: IEEEtran.bst: No hyphenation pattern has been}%
\typeout{** loaded for the language `#1'. Using the pattern for}%
\typeout{** the default language instead.}%
\else
\language=\csname l@#1\endcsname
\fi
#2}}
\providecommand{\BIBdecl}{\relax}
\BIBdecl

\bibitem{Dennards}
R.~H. Dennard, F.~H. Gaensslen, V.~L. Rideout, E.~Bassous, and A.~R. LeBlanc,
  ``Design of ion-implanted {MOSFET}'s with very small physical dimensions,''
  \emph{IEEE Journal of Solid-State Circuits}, vol.~9, no.~5, pp. 256--268,
  Oct. 1974.

\bibitem{DarkSilicon}
H.~{Esmaeilzadeh}, E.~{Blem}, R.~{St. Amant}, K.~{Sankaralingam}, and
  D.~{Burger}, ``Dark silicon and the end of multicore scaling,'' \emph{IEEE
  Micro}, vol.~32, no.~3, pp. 122--134, May 2012.

\bibitem{LRZ}
\BIBentryALTinterwordspacing
H.~Shoukourian, T.~Wilde, H.~Huber, and A.~Bode, ``Analysis of the efficiency
  characteristics of the first high-temperature direct liquid cooled petascale
  supercomputer and its cooling infrastructure,'' \emph{Journal of Parallel and
  Distributed Computing}, vol. 107, pp. 87 -- 100, 2017. [Online]. Available:
  \url{http://www.sciencedirect.com/science/article/pii/S0743731517301272}
\BIBentrySTDinterwordspacing

\bibitem{TII_conficoni}
C.~{Conficoni}, A.~{Bartolini}, A.~{Tilli}, C.~{Cavazzoni}, and L.~{Benini},
  ``Integrated energy-aware management of supercomputer hybrid cooling
  systems,'' \emph{IEEE Transactions on Industrial Informatics}, vol.~12,
  no.~4, pp. 1299--1311, Aug 2016.

\bibitem{TSC_conficoni}
C.~Conficoni, A.~Bartolini, A.~Tilli, C.~Cavazzoni, and L.~Benini, ``Hpc
  cooling: A flexible modeling tool for effective design and management,''
  \emph{IEEE Transactions on Sustainable Computing}, pp. 1--1, 2018.

\bibitem{Top500}
J.~J. Dongarra, H.~W. Meuer, E.~Strohmaier \emph{et~al.}, ``Top500
  supercomputer sites,'' [Online]; \url{https://www.top500.org/lists}, 2019,
  accessed 29 March 2019.

\bibitem{Green500}
W.-c. Feng and K.~Cameron, ``The green500 list: Encouraging sustainable
  supercomputing,'' vol.~40, no.~12.\hskip 1em plus 0.5em minus 0.4em\relax
  IEEE, 2007.

\bibitem{hackenberg2015energy}
D.~{Hackenberg}, R.~{Schöne}, T.~{Ilsche}, D.~{Molka}, J.~{Schuchart}, and
  R.~{Geyer}, ``An energy efficiency feature survey of the intel haswell
  processor,'' in \emph{2015 IEEE International Parallel and Distributed
  Processing Symposium Workshop}, May 2015, pp. 896--904.

\bibitem{OCC_ISC17}
T.~Rosedahl, M.~Broyles, C.~Lefurgy, B.~Christensen, and W.~Feng,
  ``Power/performance controlling techniques in openpower,'' in \emph{High
  Performance Computing}, J.~M. Kunkel, R.~Yokota, M.~Taufer, and J.~Shalf,
  Eds.\hskip 1em plus 0.5em minus 0.4em\relax Cham: Springer International
  Publishing, 2017, pp. 275--289.

\bibitem{cesarini2018energy}
B.~A. CESARINI, Daniele and L.~BENINI, ``Energy saving and thermal management
  opportunities in a workload-aware mpi runtime for a scientific hpc computing
  node,'' \emph{Parallel Computing is Everywhere}, vol.~32, p. 277, 2018.

\bibitem{ACPI}
``{Advanced Configuration and Power Interface (ACPI) Specification},''
  [Online]; \url{http://www.acpi.info/spec.htm}, 2019, accessed 29 March 2019.

\bibitem{fraternali_islped04}
F.~Fraternali, A.~Bartolini, C.~Cavazzoni, G.~Tecchiolli, and L.~Benini,
  ``Quantifying the impact of variability on the energy efficiency for a
  next-generation ultra-green supercomputer,'' in \emph{Proceedings of the 2014
  International Symposium on Low Power Electronics and Design}, ser. ISLPED
  '14.\hskip 1em plus 0.5em minus 0.4em\relax New York, NY, USA: ACM, 2014, pp.
  295--298.

\bibitem{lrz_lowfreq}
A.~Auweter, A.~Bode, M.~Brehm, L.~Brochard, N.~Hammer, H.~Huber, R.~Panda,
  F.~Thomas, and T.~Wilde, ``A case study of energy aware scheduling on
  supermuc,'' in \emph{Supercomputing}, J.~M. Kunkel, T.~Ludwig, and H.~W.
  Meuer, Eds.\hskip 1em plus 0.5em minus 0.4em\relax Cham: Springer
  International Publishing, 2014, pp. 394--409.

\bibitem{losalomos_sc05}
C.~Hsu and W.~Feng, ``A power-aware run-time system for high-performance
  computing,'' in \emph{SC '05: Proceedings of the 2005 ACM/IEEE Conference on
  Supercomputing}, Nov 2005, pp. 1--1.

\bibitem{GEOPM}
J.~Eastep, S.~Sylvester, C.~Cantalupo, B.~Geltz, F.~Ardanaz, A.~Al-Rawi,
  K.~Livingston, F.~Keceli, M.~Maiterth, and S.~Jana, ``Global extensible open
  power manager: A vehicle for hpc community collaboration on co-designed
  energy management solutions,'' in \emph{High Performance Computing}.\hskip
  1em plus 0.5em minus 0.4em\relax Springer International Publishing, 2017, pp.
  394--412.

\bibitem{adagio_dynamic}
B.~Rountree, D.~K. Lownenthal, B.~R. de~Supinski, M.~Schulz, V.~W. Freeh, and
  T.~Bletsch, ``Adagio: Making dvs practical for complex hpc applications,'' in
  \emph{Proceedings of the 23rd International Conference on Supercomputing},
  ser. ICS '09.\hskip 1em plus 0.5em minus 0.4em\relax New York, NY, USA: ACM,
  2009, pp. 460--469.

\bibitem{Schulz_IPDPS10}
D.~{Li}, B.~R. {de Supinski}, M.~{Schulz}, K.~{Cameron}, and D.~S.
  {Nikolopoulos}, ``Hybrid mpi/openmp power-aware computing,'' in \emph{2010
  IEEE International Symposium on Parallel Distributed Processing (IPDPS)},
  April 2010, pp. 1--12.

\bibitem{fraternali_islped14}
\BIBentryALTinterwordspacing
F.~Fraternali, A.~Bartolini, C.~Cavazzoni, G.~Tecchiolli, and L.~Benini,
  ``Quantifying the impact of variability on the energy efficiency for a
  next-generation ultra-green supercomputer,'' in \emph{International Symposium
  on Low Power Electronics and Design, ISLPED'14, La Jolla, CA, {USA} - August
  11 - 13, 2014}, 2014, pp. 295--298. [Online]. Available:
  \url{http://doi.acm.org/10.1145/2627369.2627659}
\BIBentrySTDinterwordspacing

\bibitem{fraternali2018quantifying}
F.~Fraternali, A.~Bartolini, C.~Cavazzoni, and L.~Benini, ``Quantifying the
  impact of variability and heterogeneity on the energy efficiency for a
  next-generation ultra-green supercomputer,'' \emph{IEEE Transactions on
  Parallel and Distributed Systems}, vol.~29, no.~7, pp. 1575--1588, 2018.

\bibitem{borghesi2018pricing}
\BIBentryALTinterwordspacing
A.~Borghesi, A.~Bartolini, M.~Milano, and L.~Benini, ``Pricing schemes for
  energy-efficient hpc systems: Design and exploration,'' \emph{The
  International Journal of High Performance Computing Applications}, vol.~0,
  no.~0, p. 1094342018814593, 0. [Online]. Available:
  \url{https://doi.org/10.1177/1094342018814593}
\BIBentrySTDinterwordspacing

\bibitem{borghesi2018scheduling}
A.~Borghesi, A.~Bartolini, M.~Lombardi, M.~Milano, and L.~Benini,
  ``Scheduling-based power capping in high performance computing systems,''
  \emph{Sustainable Computing: Informatics and Systems}, vol.~19, pp. 1--13,
  2018.

\bibitem{freeh2008just}
N.~{Kappiah}, V.~W. {Freeh}, and D.~K. {Lowenthal}, ``Just-in-time dynamic
  voltage scaling: Exploiting inter-node slack to save energy in mpi
  programs,'' in \emph{SC '05: Proceedings of the 2005 ACM/IEEE Conference on
  Supercomputing}, Nov 2005, pp. 33--33.

\bibitem{rapl}
\BIBentryALTinterwordspacing
H.~David, E.~Gorbatov, U.~R. Hanebutte, R.~Khanna, and C.~Le, ``Rapl: Memory
  power estimation and capping,'' in \emph{Proceedings of the 16th ACM/IEEE
  International Symposium on Low Power Electronics and Design}, ser. ISLPED
  '10.\hskip 1em plus 0.5em minus 0.4em\relax New York, NY, USA: ACM, 2010, pp.
  189--194. [Online]. Available:
  \url{http://doi.acm.org/10.1145/1840845.1840883}
\BIBentrySTDinterwordspacing

\bibitem{EEHPCJSRM}
M.~Maiterth, G.~Koenig, K.~Pedretti, S.~Jana, N.~Bates, A.~Borghesi,
  D.~Montoya, A.~Bartolini, and M.~Puzovic, ``Energy and power aware job
  scheduling and resource management: Global survey—initial analysis,'' in
  \emph{2018 IEEE International Parallel and Distributed Processing Symposium
  Workshops (IPDPSW)}.\hskip 1em plus 0.5em minus 0.4em\relax IEEE, 2018, pp.
  685--693.

\bibitem{adagio_static}
B.~Rountree, D.~K. Lowenthal, S.~Funk, V.~W. Freeh, B.~R. de~Supinski, and
  M.~Schulz, ``Bounding energy consumption in large-scale mpi programs,'' in
  \emph{Proceedings of the 2007 ACM/IEEE Conference on Supercomputing}, ser. SC
  '07.\hskip 1em plus 0.5em minus 0.4em\relax New York, NY, USA: ACM, 2007, pp.
  49:1--49:9.

\bibitem{lim2006adaptive}
M.~Y. {Lim}, V.~W. {Freeh}, and D.~K. {Lowenthal}, ``Adaptive, transparent
  frequency and voltage scaling of communication phases in {MPI} programs,'' in
  \emph{SC '06: Proceedings of the 2006 ACM/IEEE Conference on Supercomputing},
  Nov 2006, pp. 14--14.

\bibitem{kerbyson2011energy}
D.~J. Kerbyson, A.~Vishnu, and K.~J. Barker, ``Energy templates: Exploiting
  application information to save energy,'' in \emph{2011 IEEE International
  Conference on Cluster Computing}.\hskip 1em plus 0.5em minus 0.4em\relax
  IEEE, 2011, pp. 225--233.

\bibitem{simil_adagio}
S.~{Bhalachandra}, A.~{Porterfield}, S.~L. {Olivier}, and J.~F. {Prins}, ``An
  adaptive core-specific runtime for energy efficiency,'' in \emph{2017 IEEE
  International Parallel and Distributed Processing Symposium (IPDPS)}, May
  2017, pp. 947--956.

\bibitem{MVAPICH2-EA}
A.~{Venkatesh}, A.~{Vishnu}, K.~{Hamidouche}, N.~{Tallent}, D.~{Panda},
  D.~{Kerbyson}, and A.~{Hoisie}, ``A case for application-oblivious
  energy-efficient {MPI} runtime,'' in \emph{SC '15: Proceedings of the
  International Conference for High Performance Computing, Networking, Storage
  and Analysis}, Nov 2015, pp. 1--12.

\bibitem{cesarini_andare_18}
D.~Cesarini, A.~Bartolini, P.~Bonf\`{a}, C.~Cavazzoni, and L.~Benini,
  ``Countdown: A run-time library for application-agnostic energy saving in mpi
  communication primitives,'' in \emph{Proceedings of the 2Nd Workshop on
  AutotuniNg and aDaptivity AppRoaches for Energy Efficient HPC Systems}, ser.
  ANDARE '18.\hskip 1em plus 0.5em minus 0.4em\relax New York, NY, USA: ACM,
  2018, pp. 2:1--2:6.

\bibitem{cesarini_countdown}
\BIBentryALTinterwordspacing
D.~Cesarini, A.~Bartolini, P.~Bonf{\`{a}}, C.~Cavazzoni, and L.~Benini,
  ``{COUNTDOWN} - three, two, one, low power! {A} run-time library for energy
  saving in {MPI} communication primitives,'' \emph{CoRR}, vol. abs/1806.07258,
  2018. [Online]. Available: \url{http://arxiv.org/abs/1806.07258}
\BIBentrySTDinterwordspacing

\bibitem{nas}
\BIBentryALTinterwordspacing
D.~H. Bailey, \emph{NAS Parallel Benchmarks}.\hskip 1em plus 0.5em minus
  0.4em\relax Boston, MA: Springer US, 2011, pp. 1254--1259. [Online].
  Available: \url{https://doi.org/10.1007/978-0-387-09766-4_133}
\BIBentrySTDinterwordspacing

\bibitem{Luisier_3}
\BIBentryALTinterwordspacing
M.~Luisier, T.~B. Boykin, G.~Klimeck, and W.~Fichtner, ``Atomistic
  nanoelectronic device engineering with sustained performances up to 1.44
  pflop/s,'' in \emph{Proceedings of 2011 International Conference for High
  Performance Computing, Networking, Storage and Analysis}, ser. SC '11.\hskip
  1em plus 0.5em minus 0.4em\relax New York, NY, USA: ACM, 2011, pp. 2:1--2:11.
  [Online]. Available: \url{http://doi.acm.org/10.1145/2063384.2063387}
\BIBentrySTDinterwordspacing

\bibitem{Luisier_4}
M.~{Calderara}, S.~{Brück}, A.~{Pedersen}, M.~H. {Bani-Hashemian},
  J.~{VandeVondele}, and M.~{Luisier}, ``Pushing back the limit of ab-initio
  quantum transport simulations on hybrid supercomputers,'' in \emph{SC '15:
  Proceedings of the International Conference for High Performance Computing,
  Networking, Storage and Analysis}, Nov 2015, pp. 1--12.

\bibitem{PMPI}
S.~Mintchev and V.~Getov, ``Pmpi: High-level message passing in fortran77 and
  c,'' in \emph{High-Performance Computing and Networking}, B.~Hertzberger and
  P.~Sloot, Eds.\hskip 1em plus 0.5em minus 0.4em\relax Berlin, Heidelberg:
  Springer Berlin Heidelberg, 1997, pp. 601--614.

\bibitem{MVAPICH2_PSTATE}
V.~Sundriyal, M.~Sosonkina, and A.~Gaenko, ``Energy efficient communications in
  quantum chemistry applications,'' \emph{Computer Science - Research and
  Development}, vol.~29, no.~2, pp. 149--158, May 2014.

\bibitem{MVAPICH2_ALL}
V.~Sundriyal and M.~Sosonkina, ``Per-call energy saving strategies in
  all-to-all communications,'' in \emph{European MPI Users' Group
  Meeting}.\hskip 1em plus 0.5em minus 0.4em\relax Springer, 2011, pp.
  188--197.

\bibitem{MVAPICH2_Gather}
V.~Sundriyal, M.~Sosonkina, and Z.~Zhang, ``Achieving energy efficiency during
  collective communications,'' \emph{Concurrency and Computation: Practice and
  Experience}, vol.~25, no.~15, pp. 2140--2156, 2013.

\bibitem{li2004thrifty}
J.~{Li}, J.~F. {Martinez}, and M.~C. {Huang}, ``The thrifty barrier:
  energy-aware synchronization in shared-memory multiprocessors,'' in
  \emph{10th International Symposium on High Performance Computer Architecture
  (HPCA'04)}, Feb 2004, pp. 14--23.

\bibitem{benini_power_survey}
L.~{Benini}, A.~{Bogliolo}, and G.~{De Micheli}, ``A survey of design
  techniques for system-level dynamic power management,'' \emph{IEEE
  Transactions on Very Large Scale Integration (VLSI) Systems}, vol.~8, no.~3,
  pp. 299--316, June 2000.

\bibitem{HASWELL}
P.~{Hammarlund}, A.~J. {Martinez}, A.~A. {Bajwa}, D.~L. {Hill}, E.~{Hallnor},
  H.~{Jiang}, M.~{Dixon}, M.~{Derr}, M.~{Hunsaker}, R.~{Kumar}, R.~B.
  {Osborne}, R.~{Rajwar}, R.~{Singhal}, R.~{D'Sa}, R.~{Chappell}, S.~{Kaushik},
  S.~{Chennupaty}, S.~{Jourdan}, S.~{Gunther}, T.~{Piazza}, and T.~{Burton},
  ``Haswell: The fourth-generation intel core processor,'' \emph{IEEE Micro},
  vol.~34, no.~2, pp. 6--20, Mar 2014.

\bibitem{MSR_SAFE}
K.~Hoga and B.~Rountree, ``Github scalability-llnl/msr-safe, 2014,'' [Online];
  \url{https://github.com/LLNL/msr-safe}, 2019, accessed 29 March 2019.

\bibitem{score_p}
A.~Kn{\"u}pfer, C.~R{\"o}ssel, D.~a. Mey, S.~Biersdorff, K.~Diethelm,
  D.~Eschweiler, M.~Geimer, M.~Gerndt, D.~Lorenz, A.~Malony, W.~E. Nagel,
  Y.~Oleynik, P.~Philippen, P.~Saviankou, D.~Schmidl, S.~Shende,
  R.~Tsch{\"u}ter, M.~Wagner, B.~Wesarg, and F.~Wolf, ``Score-p: A joint
  performance measurement run-time infrastructure for periscope,scalasca, tau,
  and vampir,'' in \emph{Tools for High Performance Computing 2011}, H.~Brunst,
  M.~S. M{\"u}ller, W.~E. Nagel, and M.~M. Resch, Eds.\hskip 1em plus 0.5em
  minus 0.4em\relax Berlin, Heidelberg: Springer Berlin Heidelberg, 2012, pp.
  79--91.

\bibitem{extrae}
J.~Labarta, ``New analysis techniques in the cepba-tools environment,'' in
  \emph{Tools for High Performance Computing 2009}, M.~S. M{\"u}ller, M.~M.
  Resch, A.~Schulz, and W.~E. Nagel, Eds.\hskip 1em plus 0.5em minus
  0.4em\relax Berlin, Heidelberg: Springer Berlin Heidelberg, 2010, pp.
  125--143.

\bibitem{msr_safe_batch}
S.~{Walker} and M.~{McFadden}, ``Best practices for scalable power measurement
  and control,'' in \emph{2016 IEEE International Parallel and Distributed
  Processing Symposium Workshops (IPDPSW)}, May 2016, pp. 1122--1131.

\bibitem{fermata}
\BIBentryALTinterwordspacing
B.~Rountree, D.~K. Lowenthal, S.~Funk, V.~W. Freeh, B.~R. de~Supinski, and
  M.~Schulz, ``Bounding energy consumption in large-scale mpi programs,'' in
  \emph{Proceedings of the 2007 ACM/IEEE Conference on Supercomputing}, ser. SC
  '07.\hskip 1em plus 0.5em minus 0.4em\relax New York, NY, USA: ACM, 2007, pp.
  49:1--49:9. [Online]. Available:
  \url{http://doi.acm.org/10.1145/1362622.1362688}
\BIBentrySTDinterwordspacing

\bibitem{Luisier_1}
\BIBentryALTinterwordspacing
M.~Luisier, A.~Schenk, W.~Fichtner, and G.~Klimeck, ``Atomistic simulation of
  nanowires in the $s{p}^{3}{d}^{5}{s}^{*}$ tight-binding formalism: From
  boundary conditions to strain calculations,'' \emph{Phys. Rev. B}, vol.~74,
  p. 205323, Nov 2006. [Online]. Available:
  \url{https://link.aps.org/doi/10.1103/PhysRevB.74.205323}
\BIBentrySTDinterwordspacing

\bibitem{Luisier_2}
\BIBentryALTinterwordspacing
M.~Luisier, ``Atomistic simulation of transport phenomena in nanoelectronic
  devices,'' \emph{Chem. Soc. Rev.}, vol.~43, pp. 4357--4367, 2014. [Online].
  Available: \url{http://dx.doi.org/10.1039/C4CS00084F}
\BIBentrySTDinterwordspacing

\bibitem{Breiman:2001:RF:570181.570182}
\BIBentryALTinterwordspacing
L.~Breiman, ``Random forests,'' \emph{Machine Learning}, vol.~45, no.~1, pp.
  5--32, Oct 2001. [Online]. Available:
  \url{https://doi.org/10.1023/A:1010933404324}
\BIBentrySTDinterwordspacing

\bibitem{scikit-learn}
F.~Pedregosa, G.~Varoquaux, A.~Gramfort, V.~Michel, B.~Thirion, O.~Grisel,
  M.~Blondel, P.~Prettenhofer, R.~Weiss, V.~Dubourg, J.~Vanderplas, A.~Passos,
  D.~Cournapeau, M.~Brucher, M.~Perrot, and E.~Duchesnay, ``Scikit-learn:
  Machine learning in {P}ython,'' \emph{Journal of Machine Learning Research},
  vol.~12, pp. 2825--2830, 2011.

\bibitem{breiman1984classification}
L.~Breiman, J.~Friedman, R.~Olshen, and C.~Stone, ``Classification and
  regression trees. wadsworth \& brooks,'' \emph{Cole Statistics/Probability
  Series}, 1984.

\bibitem{strobl2007bias}
C.~Strobl, A.-L. Boulesteix, A.~Zeileis, and T.~Hothorn, ``Bias in random
  forest variable importance measures: Illustrations, sources and a solution,''
  \emph{BMC bioinformatics}, vol.~8, no.~1, p.~25, 2007.

\bibitem{strobl2008conditional}
C.~Strobl, A.-L. Boulesteix, T.~Kneib, T.~Augustin, and A.~Zeileis,
  ``Conditional variable importance for random forests,'' \emph{BMC
  bioinformatics}, vol.~9, no.~1, p. 307, 2008.

\bibitem{parr2018beware}
T.~Parr, K.~Turgutlu, C.~Csiszar, and J.~Howard, ``Beware default random forest
  importances,'' 2018.

\end{thebibliography}

\vspace{-1.50cm}

\begin{IEEEbiography}[{\includegraphics[width=1in, height=1.25in, clip, keepaspectratio]{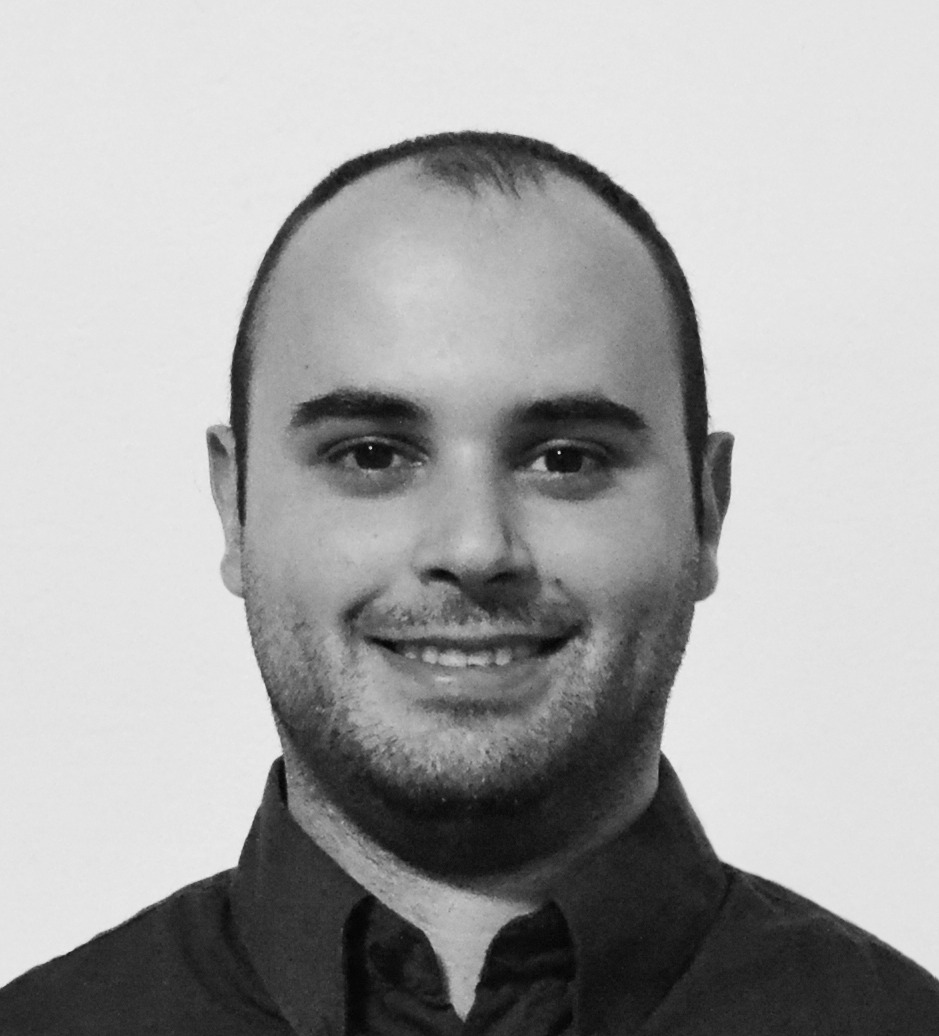}}]{Daniele Cesarini}
received a Ph.D. degree in Electrical Engineering from the University of Bologna, Italy, in 2019, where he is currently a Post-Doctoral researcher in the Department of Electrical, Electronic and Information Engineering (DEI). His research interests concern the development of SW-HW codesign strategies as well as algorithms for parallel programming support for energy-efficient HPC systems.
\end{IEEEbiography}

\vspace{-1.50cm}

\begin{IEEEbiography}[{\includegraphics[width=1in, height=1.25in, clip, keepaspectratio]{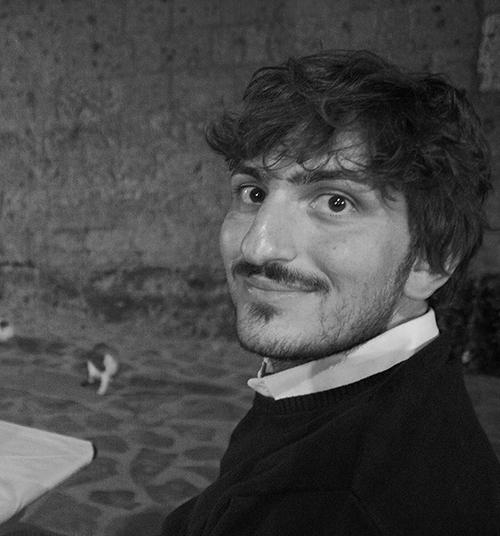}}]{Andrea Bartolini} 
received a Ph.D. degree in Electrical Engineering from the University of Bologna, Italy, in 2011. He is currently Assistant Professor in the Department of Electrical, Electronic and Information Engineering (DEI) at the University of Bologna. Before, he was Post-Doctoral researcher in the Integrated Systems Laboratory at ETH Zurich. Since 2007 Dr. Bartolini has published more than 80 papers in peer-reviewed international journals and conferences with focus on dynamic resource management for embedded and HPC systems. 
\end{IEEEbiography}

\vspace{-1.50cm}

\begin{IEEEbiography}[{\includegraphics[width=1in, height=1.25in, clip, keepaspectratio]{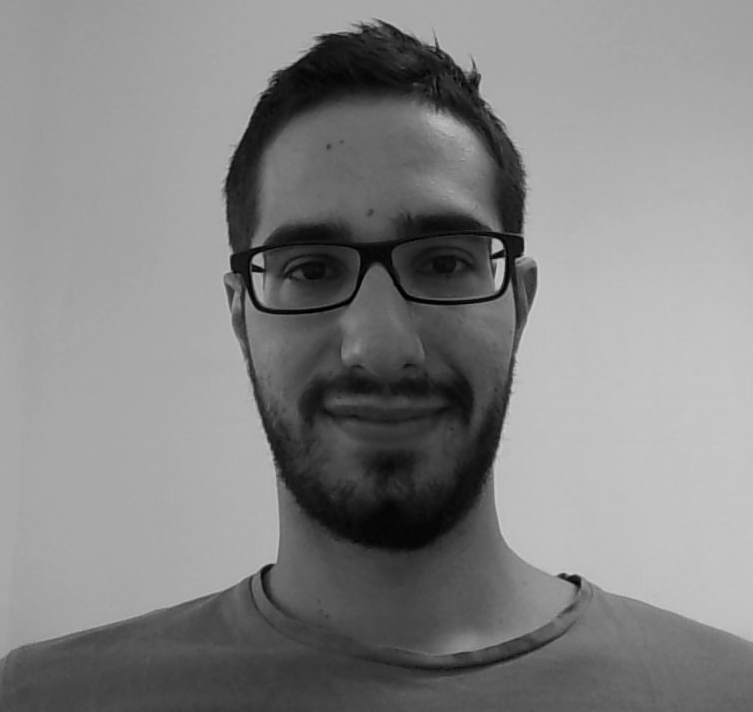}}]{Andrea Borghesi} 
is currently a Post-Doctoral researcher and Adjunct Professor at the Department of Computer Science and Engineering (DISI) of the University of Bologna, Italy. His research interests range broadly in the area of Artificial Intelligence, including Optimization, Power-awareness and Anomaly Detection in HPC systems, Scheduling and Allocation, Transprecision Computing, Machine and Deep Learning and Constraint Programming.
\end{IEEEbiography}

\vspace{-1.50cm}

\begin{IEEEbiography}[{\includegraphics[width=1in, height=1.25in, clip, keepaspectratio]{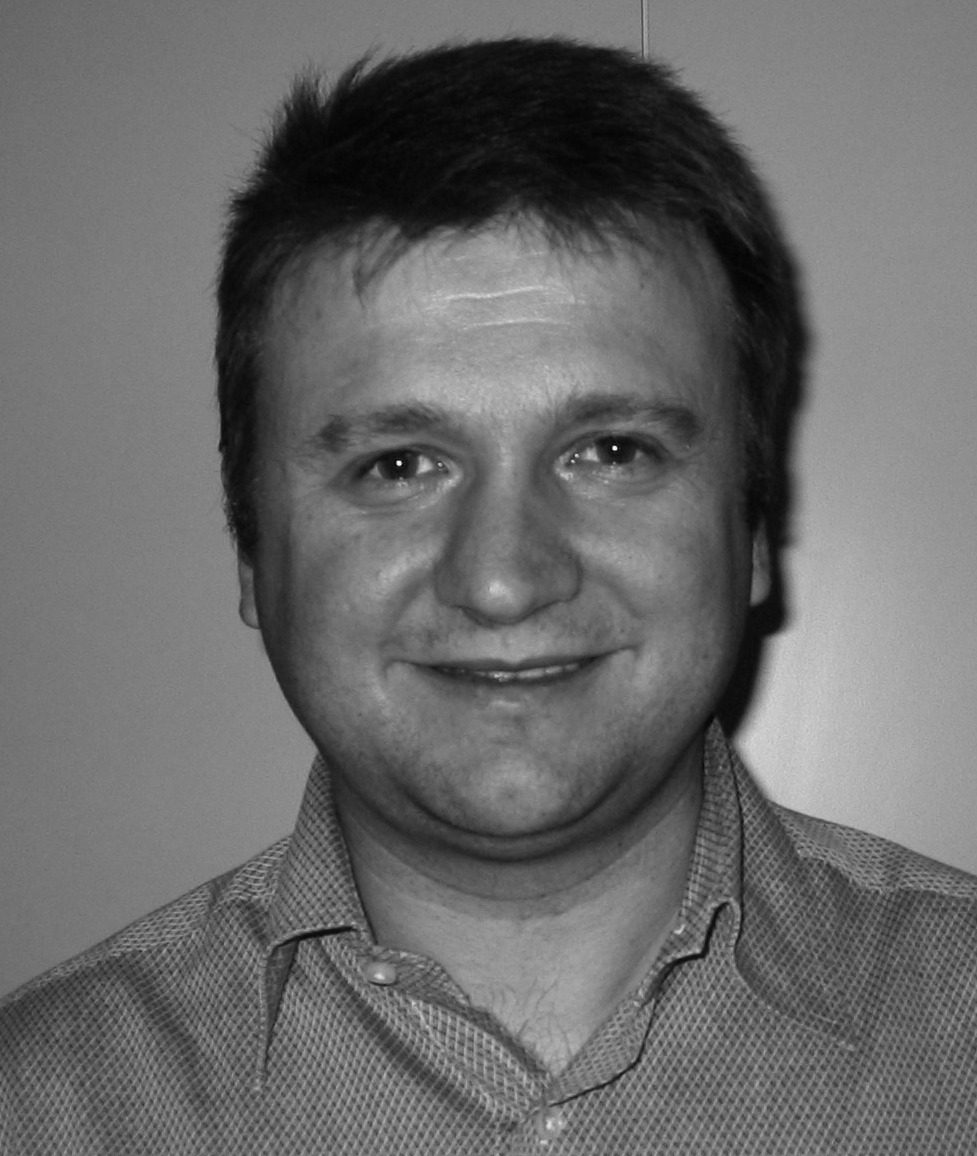}}]{Carlo Cavazzoni}
graduated cum laude in Physics from the University of Modena and earned his PhD Material Science at the International School for Advanced Studies of Trieste in 1998. 
He has authored or co-authored several papers published in prestigious international review including Science, Physical Review Letters, Nature Materials. Currently in the HPC Business Unit of CINECA, he is responsible for the R\&D, HPC infrastructure evolution and collaborations with scientific communities. 
\end{IEEEbiography}

\vspace{-1.50cm}

\begin{IEEEbiography}[{\includegraphics[width=1in, height=1.25in, clip, keepaspectratio]{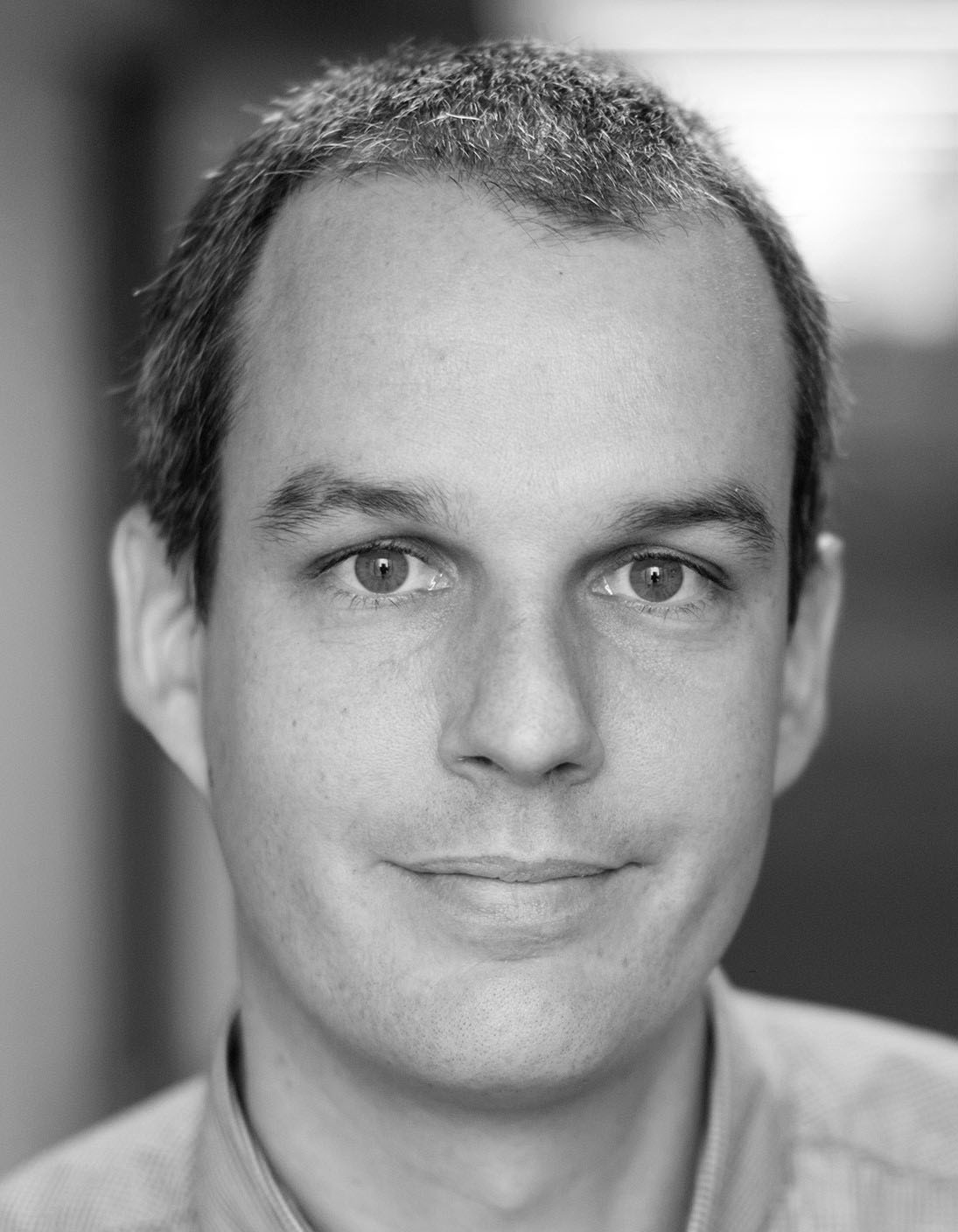}}]{Mathieu Luisier}
is Associate Professor of Computational Nanoelectronics at ETH Zurich (Switzerland). His research focuses on the development of advanced technology computer aided design (TCAD) tools and their application to modern nano-devices such as next-generation transistors or non-volatile random access memory cells. He has published more than 200 articles in peer-reviewed journals and conferences and is a member of IEEE.
\end{IEEEbiography}

\vspace{-1.50cm}

\begin{IEEEbiography}[{\includegraphics[width=1in, height=1.25in, clip, keepaspectratio]{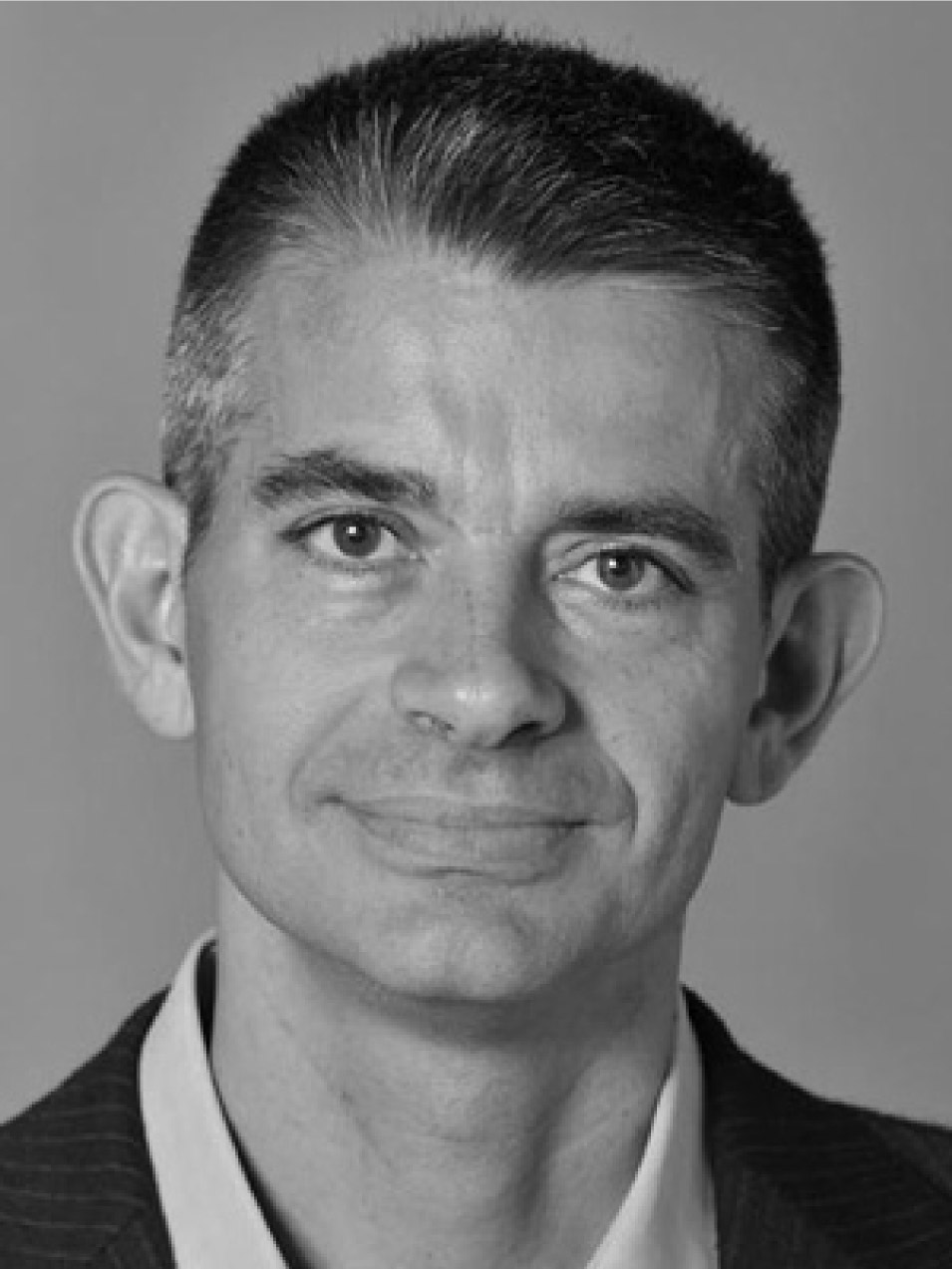}}]{Luca Benini} 
is professor of Digital Circuits and Systems at ETH Zurich, Switzerland, and is also professor at University of Bologna, Italy. His research interests are in system design of energy-efficient multicore SoC, smart sensors and sensor networks. He has published more than 800 papers in peer reviewed international journals and conferences, four books and several book chapters. He is a fellow of the ACM and Member of the Academia Europea. He is the recipient of the IEEE CAS Mac Van Valkenburg Award 2016.

\end{IEEEbiography}

\end{document}